\documentclass{raa}
\usepackage{longtable}
\usepackage{graphicx,times}             
\usepackage{natbib}
\usepackage{amssymb,amsmath}
\bibpunct{(}{)}{;}{a}{}{,}
\usepackage[pagebackref=true]{hyperref}
\begin{document}
\defcitealias{ngc5904}{Paper I}
\defcitealias{ngc6205}{Paper II}
\defcitealias{ngc288}{Paper III}
\defcitealias{ngc6362}{Paper IV}
\defcitealias{ngc6397}{Paper V}
\defcitealias{nardiello2018}{NLP18}
\defcitealias{stetson2019}{SPZ19}
\defcitealias{grundahl1999}{GCL99}
\defcitealias{vasiliev2021}{VB21}
\defcitealias{baumgardt2021}{BV21}
\defcitealias{ccm89}{CCM89}
\defcitealias{bonatto2013}{BCK13}
\defcitealias{sfd98}{SFD98}

\title{Isochrone Fitting of Galactic Globular Clusters -- VI. High-latitude Clusters NGC\,5024 (M53), NGC\,5053, NGC\,5272 (M3), NGC\,5466, and NGC\,7099 (M30)
}
   \volnopage{Vol.0 (20xx) No.0, 000--000}      
   \setcounter{page}{1}          

\author{G.~A.~Gontcharov
\inst{1}
\and S.~S.~Savchenko
\inst{1,2,3}
\and A.~A.~Marchuk
\inst{1,2}
\and C.~J.~Bonatto
\inst{4}
\and O.~S.~Ryutina
\inst{2}
\and M.~Yu.~Khovritchev
\inst{1,2}
\and V.~B.~Il'in
\inst{1,2,5}
\and A.~V.~Mosenkov
\inst{6}
\and D.~M.~Poliakov
\inst{1}
\and A.~A.~Smirnov
\inst{1}
}
\institute{Central (Pulkovo) Astronomical Observatory, Russian Academy of Sciences, Pulkovskoye chaussee 65/1, St. Petersburg 196140, Russia; {\it georgegontcharov@yahoo.com}\\
\and
Saint Petersburg State University, Universitetskij pr. 28, St. Petersburg 198504, Russia\\
\and
Special Astrophysical Observatory, Russian Academy of Sciences, 369167 Nizhnij Arkhyz, Russia\\
\and
Departamento de Astronomia, Instituto de F\'isica, UFRGS, Av. Bento Gon\c{c}alves, 9500, Porto Alegre, RS, Brazil\\
\and
Saint Petersburg University of Aerospace Instrumentation, Bol. Morskaya ul. 67A, St. Petersburg 190000, Russia\\
\and
Department of Physics and Astronomy, Brigham Young University, N283 ESC, Provo, UT 84602, USA\\
\vs\no
   {\small Received 20xx month day; accepted 20xx month day}}

\abstract{We fit various colour--magnitude diagrams (CMDs) of the high-latitude Galactic globular clusters NGC\,5024 (M53), NGC\,5053, NGC\,5272 (M3), NGC\,5466, and NGC\,7099 (M30)
by isochrones from the Dartmouth Stellar Evolution Database and Bag of Stellar Tracks and Isochrones for $\alpha$--enrichment [$\alpha$/Fe]$=+0.4$.
For the CMDs, we use data sets from {\it Hubble Space Telescope}, {\it Gaia}, and other sources utilizing, at least, 25 photometric filters for each cluster.
We obtain the following characteristics with their statistic uncertainties for NGC\,5024, NGC\,5053, NGC\,5272, NGC\,5466, and NGC\,7099, respectively:
metallicities [Fe/H]$=-1.93\pm0.02$, $-2.08\pm0.03$, $-1.60\pm0.02$, $-1.95\pm0.02$, and $-2.07\pm0.04$ dex with their systematic uncertainty 0.1 dex;
ages $13.00\pm0.11$, $12.70\pm0.11$, $11.63\pm0.07$, $12.15\pm0.11$, and $12.80\pm0.17$ Gyr with their systematic uncertainty 0.8 Gyr;
distances (systematic uncertainty added) $18.22\pm0.06\pm0.60$, $16.99\pm0.06\pm0.56$, $10.08\pm0.04\pm0.33$, $15.59\pm0.03\pm0.51$, and $8.29\pm0.03\pm0.27$ kpc;
reddenings $E(B-V)=0.023\pm0.004$, $0.017\pm0.004$, $0.023\pm0.004$, $0.023\pm0.003$, and $0.045\pm0.002$ mag  with their systematic uncertainty 0.01 mag;
extinctions $A_\mathrm{V}=0.08\pm0.01$, $0.06\pm0.01$, $0.08\pm0.01$, $0.08\pm0.01$, and $0.16\pm0.01$ mag  with their systematic uncertainty 0.03 mag,
which suggest the total Galactic extinction $A_\mathrm{V}=0.08$ across the whole Galactic dust to extragalactic objects at the North Galactic pole.
The horizontal branch morphology difference of these clusters is explained by their different metallicity, age, mass-loss efficiency, 
and loss of low-mass members in the evolution of the core-collapse cluster NGC\,7099 and loose clusters NGC\,5053 and NGC\,5466.
\keywords{Hertzsprung--Russell and colour--magnitude diagrams --
dust, extinction --
globular clusters: general --
globular clusters: individual: NGC\,5024, NGC\,5053, NGC\,5272, NGC\,5466, NGC\,7099}
}

\authorrunning{G. A. Gontcharov et al.}            
\titlerunning{Isochrone fitting of Galactic globular clusters (VI)}  

\maketitle

\section{Introduction}
\label{intro}

In \citet[][hereafter Paper I]{ngc5904}, \citet[][hereafter Paper II]{ngc6205}, \citet[][hereafter Paper III]{ngc288}, \citet[][hereafter Paper IV]{ngc6362}, and
\citet[][hereafter Paper V]{ngc6397} 
we estimated important parameters (interstellar extinction in many filters, metallicity [Fe/H], age, and distance $R$ from the Sun) for several Galactic globular clusters (GCs)
via fitting their colour--magnitude diagrams (CMDs) by theoretical isochrones based on stellar evolution models.
Each CMD can provide us with independent estimates of [Fe/H], age, $R$, and reddening corresponding to the colour of this CMD.
The novelty of our results is due to recent appearances and improvements for models/isochrones and photometric data sets of individual cluster members in ultraviolet (UV), optical, 
and infrared (IR) bands, with accurate selection of the members by use of the precise parallaxes and proper motions (PMs) from the 
{\it Hubble Space Telescope (HST}; \citealt{libralato2022}) and {\it Gaia} Data Release 3 (DR3; \citealt{gaiadr3}).
Both the recent data sets and isochrones successfully reproduce the main stages of stellar evolution, namely the main sequence (MS), turn-off (TO), subgiant branch (SGB), 
red giant branch (RGB), horizontal branch (HB), and asymptotic giant branch (AGB).

\begin{table*}
\def\baselinestretch{1}\normalsize\scriptsize
\caption[]{Some Properties of the Clusters under Consideration. \\
$r_t/r_c$ is the ratio of tidal and core radii, 
$\Delta(V-I)$ is the median colour difference between the HB and RGB from \citet{dotter2010},
$R$ is the distance from the Sun,
$\delta Y_{2G,1G}$ is the average helium difference between the second and first stellar generations,
$\overline{\Delta E(B-V)}$ and $\Delta E(B-V)_\mathrm{max}$ are the mean and maximum differential reddening, respectively.
We use the $R$ and [Fe/H] estimates of \citet{arellano2024} for the RRc variables, with the [Fe/H] estimates being on the [Fe/H] scale of \citet{carretta2009}.
}
\label{properties}
\[
\begin{tabular}{lccccc}
\hline
\noalign{\smallskip}
Property                                                               &  NGC\,5024  &  NGC\,5053 & NGC\,5272 & NGC\,5466 & NGC\,7099 \\
\hline
\noalign{\smallskip}
RA J2000 (h~m~s) from \citet{goldsbury2010}                            & \hphantom{$-$}13 12 55 & \hphantom{$-$}13 16 27 & \hphantom{$-$}13 42 12 & \hphantom{$-$}14 05 27 & \hphantom{$-$}21 40 22 \\
Dec. J2000 ($\degr$ $\arcmin$ $\arcsec$) from \citet{goldsbury2010}    &   $+18$ 10 05          &   $+$17 42 01    &   $+$28 22 38    &   $+$28 32 04   &   $-$23 10 48    \\
Galactic longitude ($\degr$) from \citet{goldsbury2010}                &   332.9624             &   335.6983       &   42.2164        &   42.1499       &   27.1791        \\
Galactic latitude ($\degr$) from \citet{goldsbury2010}                 &   $+79.7641$           &   $+78.9461$     &   $+78.7069$     &   $+73.5923$    &   $-46.8354$     \\
Angular radius (arcmin) from \citet{bica2019}                          &   10.0                 &   5.0            &   13.0           &   5.5           &   9.0            \\
Tidal radius (arcmin) from \citet{hunt2023}                            &   6.8                  &   4.5            &   8.6            &   6.4           &   8.7            \\
Tidal radius (arcmin) from \citetalias{baumgardt2021}                  &   34.0                 &   18.0           &   43.7           &   16.4          &   28.9           \\
Truncation radius (arcmin) from this study                             &   14.5                 &   10.0           &   23.0           &   17.0          &   12.5           \\
Core density (solar mass per cubic pc) from \citetalias{baumgardt2021} &          1259          &         3        &         6026     &     8           &   5\,495\,409     \\
$r_t/r_c$ from \citetalias{baumgardt2021}                              &             98         &           10     &         132      &         12      &        791       \\
$\Delta(V-I)$ from \citet{dotter2010}                                  &       $0.851\pm0.010$  &  $0.782\pm0.041$ &  $0.736\pm0.014$ & $0.721\pm0.025$ &  $0.872\pm0.006$ \\
$\tau_{HB}$ index from \citet{torelli2019}                             & $6.67\pm0.16$          & $4.35\pm0.11$    & $4.13\pm0.05$    & $5.02\pm0.10$   & $6.40\pm0.20$  \\
HB type from \citet{torelli2019}                                       &  $+0.89\pm0.06$        & $+0.46\pm0.12$   &  $+0.21\pm0.02$  & $+0.62\pm0.11$  & $+0.90\pm0.10$    \\
HB type from \citet{arellano2024}                                      &  $+0.81$        & $+0.50$   &  $+0.08$  & $+0.58$  & $+0.89$    \\
Mean HB type from this study                                           &  $+0.83\pm0.02$        & $+0.74\pm0.17$   &  $+0.10\pm0.03$  & $+0.62\pm0.08$  & $+0.83\pm0.09$    \\
$R$ (kpc) from \citet{harris}, 2010 revision\footnotemark\             &   17.9                 &   17.4           &   10.2           & 16.0            &   8.1             \\
$R$ (kpc) from \citetalias{baumgardt2021}                              &   $18.50\pm0.18$       &  $17.54\pm0.23$  & $10.175\pm0.082$ & $16.12\pm0.16$  &   $8.458\pm0.090$ \\
$R$ (kpc) from \citet{arellano2024}                                    &   $18.0\pm0.4$         &  $16.7\pm0.4$    &  $10.0\pm0.4$    & $16.0\pm0.6$    &   $8.1$   \\
$R$ (kpc) from \citet{hunt2023}                                        &   $14.42\pm0.34$       &  $19.21\pm1.48$  &   $9.69\pm0.09$  & $17.67\pm0.73$  &   $7.83\pm0.12$   \\
$[$Fe$/$H$]$ from \citet{carretta2009}                                 &   $-2.06\pm0.09$       &  $-2.30\pm0.08$  &   $-1.50\pm0.05$ & $-2.31\pm0.09$  &   $-2.33\pm0.02$  \\
$[$Fe$/$H$]$ from \citet{meszaros2020}                                 &   $-1.89\pm0.11$       &  $-2.06\pm0.11$  &   $-1.39\pm0.13$ & $-1.83\pm0.11$  &            \\
Spectroscopic $[$Fe$/$H$]$ from \citet{jurcsik2023}                    &   $-2.03\pm0.04$       &  $-2.21\pm0.03$  &   $-1.47\pm0.02$ & $-1.92\pm0.05$  &   $-2.32\pm0.10$  \\
Photometric $[$Fe$/$H$]$ from \citet{jurcsik2023}                      &   $-1.87\pm0.02$       &  $-2.01\pm0.03$  &   $-1.48\pm0.02$ & $-1.98\pm0.02$  &   $-2.18\pm0.05$  \\
$[$Fe$/$H$]$ from \citet{arellano2024}                                 &   $-1.85\pm0.13$       &  $-2.05\pm0.18$  &   $-1.57\pm0.14$ & $-1.89\pm0.21$  &   $-2.14$  \\
$\delta Y_{2G,1G}$ from \citet{milone2018}                             &   $0.013\pm0.007$      & $-0.002\pm0.013$ &  $0.016\pm0.005$ & $0.002\pm0.017$ & $0.015\pm0.010$ \\
Age (Gyr) from \citet{dotter2010}                                      &  $13.25\pm0.50$        & $13.50\pm0.75$   & $12.50\pm0.50$   & $13.00\pm0.75$  & $13.25\pm1.00$  \\
Age (Gyr) from \citet{forbes2010}                                      &  $12.67\pm0.64$        & $12.29\pm0.51$   & $11.39\pm0.51$   & $13.57\pm0.64$  & $12.93\pm0.64$ \\
Age (Gyr) from \citet{vandenberg2013}                                  & $12.25\pm0.25$         & $12.25\pm0.38$   & $11.75\pm0.25$   & $12.50\pm0.25$  & $13.00\pm0.25$ \\
Age (Gyr) from \citet{valcin2020}       & $13.31^{+0.66}_{-0.57}$ & $13.84^{+0.50}_{-0.58}$ & $12.60^{+0.66}_{-0.66}$ & $12.31^{+0.60}_{-0.40}$ & $12.82^{+0.33}_{-0.50}$ \\
$\overline{\Delta E(B-V)}$ (mag) from \citetalias{bonatto2013}         &   $0.030\pm0.009$      &  $0.029\pm0.011$ &  $0.031\pm0.009$ & $0.024\pm0.009$ &   $0.030\pm0.010$ \\
$\Delta E(B-V)_\mathrm{max}$ (mag) from \citetalias{bonatto2013}       &   0.068                & 0.058            & 0.063            & 0.048           & 0.064    \\
$E(B-V)$ (mag) from \citet{harris}, 2010 revision                      &           0.02         &       0.01       &        0.01      &       0.00      &      0.03 \\
$E(B-V)$ (mag) from \citetalias{sfd98}                                 &           0.02         &       0.02       &        0.01      &       0.02      &      0.05  \\ 
$E(B-V)$ (mag) from \citet{schlaflyfinkbeiner2011}                     &           0.02         &       0.01       &        0.01      &       0.01      &      0.04 \\ 
$E(B-V)$ (mag) from \citet{planck}                                     &           0.02         &       0.02       &        0.02      &       0.02      &      0.06 \\
$E(B-V)$ (mag) from \citet{lallement2019}                              &           0.01         &       0.01       &        0.01      &       0.01      &      0.02  \\
$E(B-V)$ (mag) from \citet{green2019}                                  &           0.02         &       0.01       &        0.07      &       0.03      &      0.06 \\
$E(B-V)$ (mag) from \citet{gmk2023}                                    &           0.04         &       0.05       &        0.02      &       0.05      &      0.07 \\
\hline
\end{tabular}
\]
\end{table*}
\footnotetext{The commonly used database of GCs by \citet{harris} (\url{https://www.physics.mcmaster.ca/~harris/mwgc.dat}), 2010 revision.}

\begin{table*}
\def\baselinestretch{1}\normalsize\small
\caption{Some Previous Mutual Estimates of the Parameters of the Clusters under Consideration. \\
We average results obtained by different methods inside each study.
Age is in Gyr, $R$ is in kpc.
}
\label{diversity}
\[
\begin{tabular}{lllll}
\hline
\noalign{\smallskip}
Study                   &  [Fe/H]          &   Age    & $R$                 & $E(B-V)$ or $A_\mathrm{V}$ \\
\hline
\noalign{\smallskip}
        \multicolumn{5}{c}{NGC\,5024} \\
\citet{dotter2010}      & $-2.00$          & $13.25\pm0.50$ & $18.95$                  & $E(B-V)=0.023$ \\
\hline
\noalign{\smallskip}
        \multicolumn{5}{c}{NGC\,5053} \\
\citet{dotter2010}      & $-2.40$          & $13.50\pm0.75$ & $18.04$                  & $E(B-V)=0.021$ \\
\citet{arellano2010}    & $-1.97\pm0.16$   & $12.5\pm2.0$   & $16.7\pm0.3$             & \\
\citet{paust2010}       & $-1.99$          & $12.0$         & $17.27$                  & $E(B-V)=0.04$ \\
\citet{nikitha2022}     & $-1.9$           & $12.5\pm2.0$   & $17.06$                  & \\
\hline
\noalign{\smallskip}
        \multicolumn{5}{c}{NGC\,5272} \\
\citet{dotter2010}      & $-1.60$          & $12.50\pm0.50$ & $10.22$                  & $E(B-V)=0.018$ \\
\citet{paust2010}       & $-1.57$          & $12.0$         & $10.17$                  & $E(B-V)=0.04$ \\
\citet{stenning2016}    & $-1.465\pm0.003$ & $11.80\pm0.06$ & $10.20\pm0.03$           & $A_\mathrm{V}=0.075\pm0.002$ \\
\citet{denissenkov2017} & $-1.55$          & $12.6$         & $9.91$                   & $E(B-V)=0.013$ \\
\citet{preetkaur2022}   & $-1.57$          & $11.56$        & $10.47$                  & $E(B-V)=0.015$ \\
\hline
\noalign{\smallskip}
        \multicolumn{5}{c}{NGC\,5466} \\
\citet{dotter2010}      & $-2.10$          & $13.00\pm0.75$ & $16.43$                  & $E(B-V)=0.023$ \\
\citet{paust2010}       & $-2.22$          & $12.0$         & $16.28$                  & $E(B-V)=0.05$ \\
\hline
\noalign{\smallskip}
        \multicolumn{5}{c}{NGC\,7099} \\
\citet{dotter2010}      & $-2.40$          & $13.25\pm1.00$ &  $8.80$                  & $E(B-V)=0.053$ \\
\citet{paust2010}       & $-2.12$          & $13.0$         &  $8.14$                  & $E(B-V)=0.07$ \\
\citet{kains2013}       & $-2.06\pm0.12$   & $13.0\pm1.0$   & $8.33\pm0.45$            &  \\
\hline
\end{tabular}
\]
\end{table*}

In this paper, we fit five high-latitude clusters NGC\,5024 (Messier 53, M53), NGC\,5053, NGC\,5272 (Messier 3, M3), NGC\,5466, and NGC\,7099 (Messier 30, M30).
Some properties of these clusters are presented in Table~\ref{properties}.
It shows that tidal radius, metallicity, distance, and age of these clusters are quite uncertain and, hence, should be clarified.
This uncertainty is evident from all previous studies [see a list of them in \citet[][hereafter BV21]{baumgardt2021}].
Some previous studies with mutual estimates of several parameters are presented in Table~\ref{diversity}.
The same method of isochrone fitting, even when applied to the same data sets for decades, has yielded dissimilar estimates. 
This inconsistency can be attributed to significant reasons: the large uncertainty in the photometry used, which is at a level of 0.1 mag, and the contamination of the data by 
non-members. 
Both issues can be addressed with recent data sets. 
These data sets offer photometry with precision to a few hundredths of a magnitude and are complemented by highly accurate proper motions and parallaxes. 
This study employs such advanced data sets to overcome the aforementioned challenges.

The clusters under consideration have a lot of estimates of metallicity [Fe/H].
Four of the clusters with a low metallicity [Fe/H]$\approx-2$ (i.e. all but NGC\,5272)
are among the most metal-poor GCs and, hence, they are our first test of the models/isochrones in such a low-metallicity regime.
Table~\ref{properties} shows examples of large discrepancies as between spectroscopic and photometric estimates of [Fe/H], as between various spectroscopic estimates themselves.
A detailed study of this issue is presented by \citet{mucciarelli2020}. They argue in favour of photometrically and against spectroscopically derived [Fe/H] for low-metallicity GCs.
Since the slopes of the RGB and faint MS are sensitive to [Fe/H], our isochrone fitting of CMDs with well-populated bright RGB or faint MS can provide new independent [Fe/H] estimates 
for these clusters. Note that we estimate [Fe/H] separately for each model.
As noted in \citetalias{ngc6397}, crowding or poor astrometry at the cluster centres, saturation and completeness effects, photometric errors and 
helium abundance $Y$ uncertainty may result in an uncertainty of about 0.2 dex in our [Fe/H] estimate obtained from the pair of a CMD and a model.

Uncertain [Fe/H], age, and distance of the four low-metallicity clusters do not allow one to make a final conclusion about similarity and origin of them.
\citet{yoon2002} find that these clusters display a planar alignment of their positions and orbits in the outer halo, which, being combined with other of their properties, 
suggests their captured origin from a satellite galaxy.
This is supported by \citet{chun2020} and should be confirmed, at least, for NGC\,5024 and NGC\,5053, whose proximity in space to each other cannot be accidental.

Even foreground reddening of all the clusters [e.g. reddening $E(B-V)$], certainly being very low, is not well defined, as suggested by a noticeable differential reddening (DR) found by 
\citet[][hereafter BCK13]{bonatto2013} and presented in Table~\ref{properties}.
In this context, these clusters are interesting targets for further research.
In particular, these clusters can be probes of total Galactic extinction at high latitudes, while very low reddening and extinction makes them a test of the models/isochrones 
in such a low-extinction regime, where they should not predict an unreal negative extinction/reddening.

Table~\ref{properties} also shows that these clusters differ in the HB morphology indexes: $\Delta(V-I)$ defined by \citet{dotter2010}, $\tau_{HB}$ defined by
\citet{torelli2019}, and HB type\footnote{The HB type is defined as $(N_B-N_R)/(N_B+N_V+N_R)$, where $N_B$, $N_V$, and $N_R$ are the number of stars that lie blueward of the 
instability strip, the number of RR~Lyrae variables, and the number of stars that lie redward of the instability strip, respectively \citep{lee1994}.}
calculated by \citet{torelli2019} and \citet{arellano2024}, albeit similar, at least, for NGC\,5024 and NGC\,7099.
This forces us to analyse the HB morphology difference of the clusters in Sect.~\ref{hb}.
This difference may be related to the difference in structural parameters of these clusters presented in Table~\ref{properties}:
e.g. a core-collapse NGC\,7099 with a compact core differs strikingly from loose NGC\,5466 and NGC\,5053 in core density and ratio $r_t/r_c$ of tidal and core radius.

The clusters under consideration have, at least, two stellar generations \citep{milone2017} with a similar $\alpha$--enrichment [$\alpha$/Fe]$\approx0.4$ 
\citep{carretta2010,kacharov2015,boberg2015,boberg2016,masseron2019,chun2020}, 
but slightly different helium enrichment, as seen from Table~\ref{properties}.
As in our previous studies, we check that this helium enrichment is mild in order to fit a cluster dominant generation or a mix of generations with reliable results.
Specifically, we fit isochrones to all considered CMDs using a reasonable grid of helium abundance $Y$, metallicity [Fe/H], distance, reddening, and age.
We use the $\alpha$--enhanced theoretical models of stellar evolution and corresponding isochrones from
Dartmouth Stellar Evolution Database (DSED, \citealt{dotter2007,dotter2008})\footnote{\url{http://stellar.dartmouth.edu/models/}} and
a Bag of Stellar Tracks and Isochrones (BaSTI, \citealt{newbasti,pietrinferni2021})\footnote{\url{http://basti-iac.oa-abruzzo.inaf.it/index.html}}, which turned out to be suitable 
for such a fitting in our previous studies.
Also, we use the BaSTI extended set of zero-age horizontal branch (ZAHB) models with a stochastic mass loss between the MS and HB.
For control of our DSED fitting, we use the DSED HB and AGB isochrones, which exist for some filters.
The DSED isochrones for $Y=0.25$ and $0.33$ and BaSTI isochrones for $Y=0.25$ and $Y=0.275$ are used to interpolate or extrapolate isochrones for other $Y$.
Such interpolation produces a negligible uncertainty less than 0.01 mag in any CMD, since the initial isochrones are close to each other and they are presented by the same 
evolutionary points.

\begin{figure}
\includegraphics{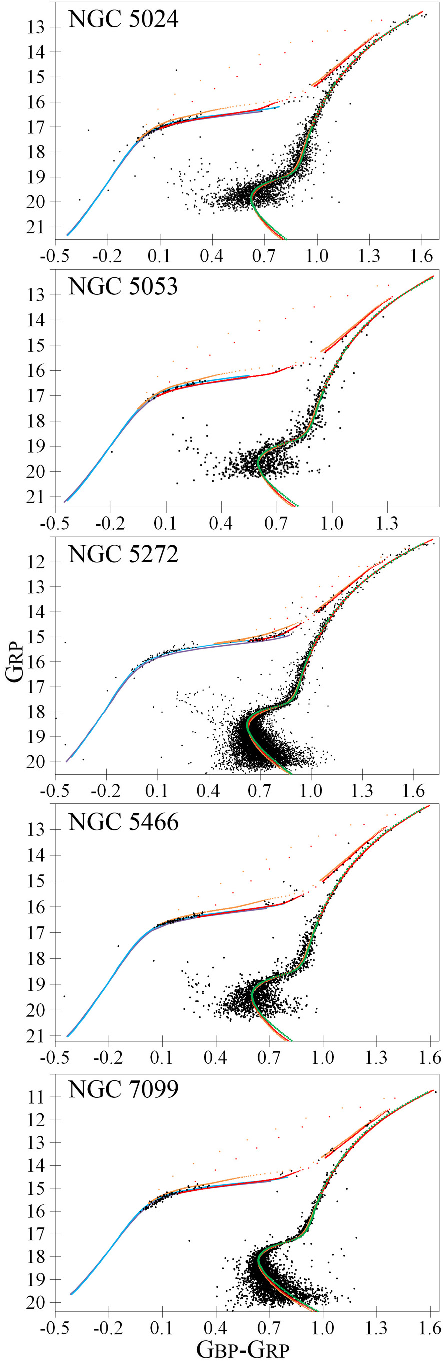}
\caption{$G_\mathrm{BP}-G_\mathrm{RP}$ versus $G_\mathrm{RP}$ CMDs for the {\it Gaia} DR3 clusters members.
The isochrones for $Y=0.25$ from BaSTI (red), BaSTI ZAHB (purple), and DSED (green), as well as for $Y=0.275$ from BaSTI (orange) and BaSTI ZAHB (blue)
are calculated with the best-fitting parameters from Table~\ref{cmds}.
RR~Lyrae variables are eliminated.
}
\label{gaia}
\end{figure}

\begin{figure}
\includegraphics{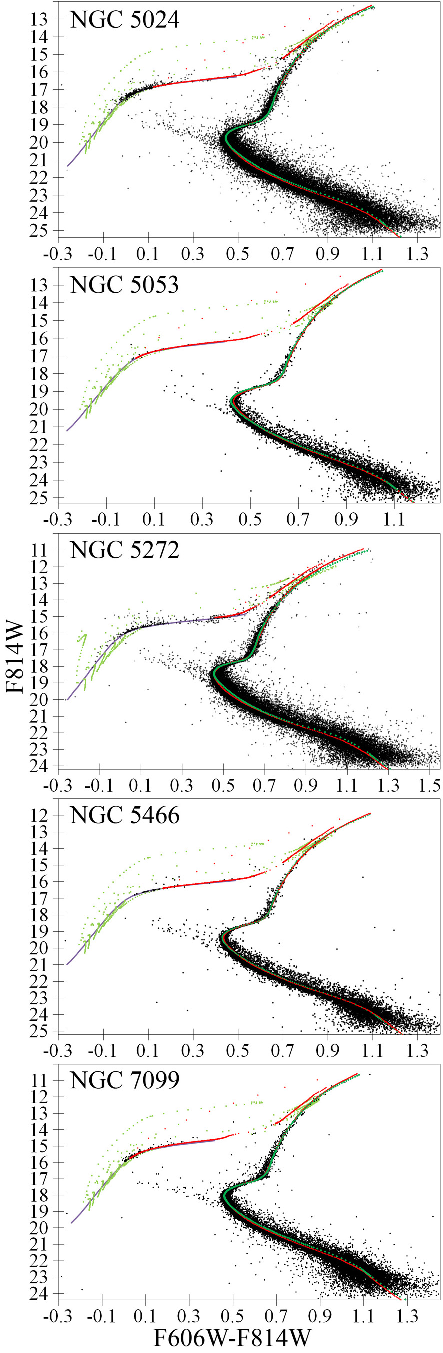}
\caption{{\it HST}/ACS $F606W-F814W$ versus $F814W$ CMDs for the \citetalias{nardiello2018} data sets.
The isochrones for $Y=0.25$ from BaSTI (red), BaSTI ZAHB (purple), DSED (green), and DSED HB/AGB (light green) are calculated with the best-fitting parameters from Table~\ref{cmds}.
RR~Lyrae variables are retained.
}
\label{hst}
\end{figure}

\begin{figure}
\includegraphics{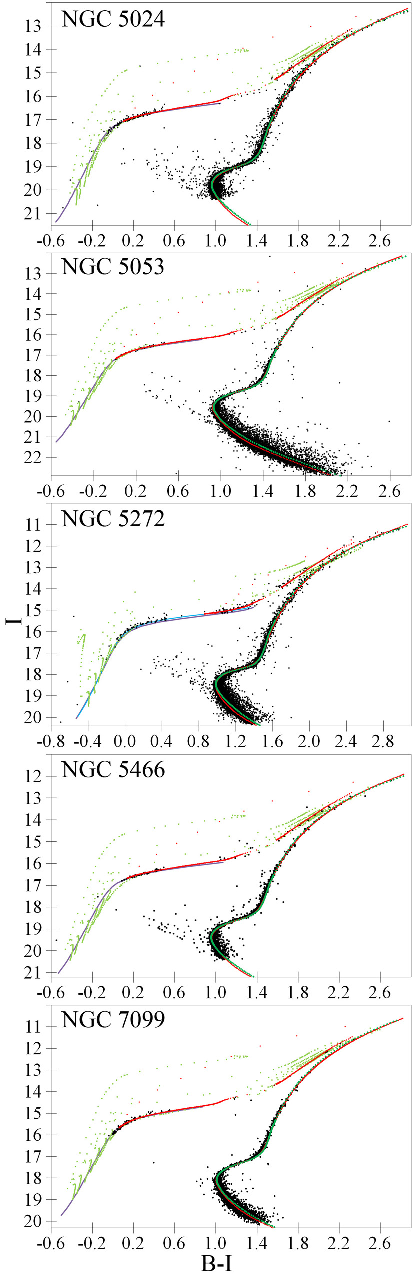}
\caption{$B-I$ versus $I$ CMDs for the {\it Gaia} cluster members from the \citetalias{stetson2019} data sets (all stars of the \citetalias{stetson2019} data set for NGC\,5053).
The isochrones for $Y=0.25$ from BaSTI (red), BaSTI ZAHB (purple), DSED (green), and DSED HB/AGB (light green), as well as the NGC\,5272 BaSTI ZAHB for $Y=0.275$ (blue) 
are calculated with the best-fitting parameters from Table~\ref{cmds}.
RR~Lyrae variables are eliminated.
}
\label{stetson}
\end{figure}

\begin{figure}
\includegraphics{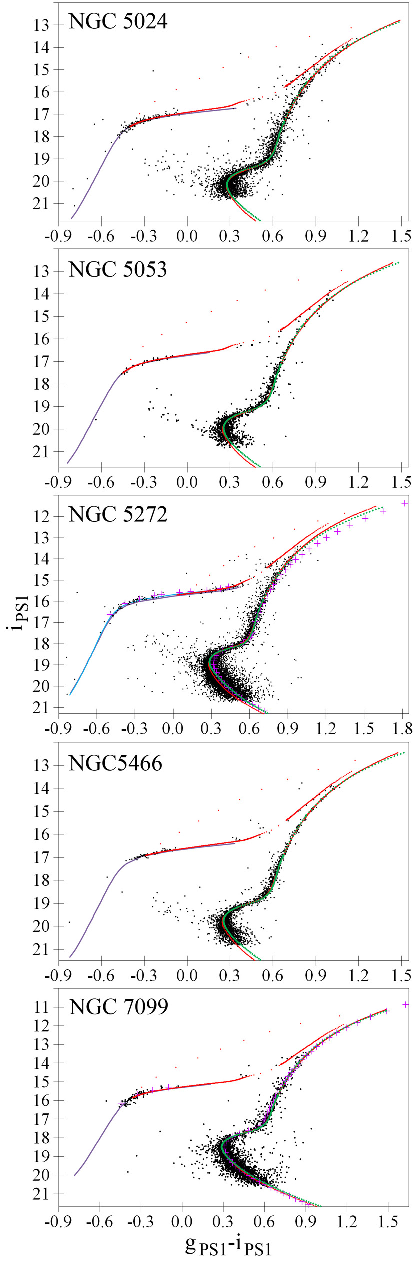}
\caption{$g_\mathrm{PS1}-i_\mathrm{PS1}$ versus $i_\mathrm{PS1}$ CMDs for the {\it Gaia} cluster members from the PS1 data sets.
The fiducial sequences of \citet{bernard2014} for NGC\,5272 and NGC\,7099 are shown as the purple crosses.
The isochrones for $Y=0.25$ from BaSTI (red), BaSTI ZAHB (purple), and DSED (green), as well as the NGC\,5272 BaSTI ZAHB for $Y=0.275$ (blue) are calculated with 
the best-fitting parameters from Table~\ref{cmds}.
RR~Lyrae variables are eliminated.
}
\label{ps1}
\end{figure}

Fig.~\ref{gaia} presents an example of CMDs where the BaSTI isochrones with different $Y$ are drawn (together with the DSED isochrones for $Y=0.25$). 
Other CMDs demonstrate a similar pattern, but we do not show the BaSTI isochrones with $Y=0.275$ in Figs~\ref{hst}--\ref{ps1} for clarity.
A noticeable separation of the isochrones with $Y=0.25$ and 0.275 at the HB and AGB allows us to conclude that most stars fit $Y=0.25$.
This agrees, for example, with a robust estimate $Y=0.252\pm0.003$ for NGC\,7099 \citep{mucciarelli2014}.
We adopt $Y=0.25$ for all domains of our CMDs, except the HB bluer than the RR~Lyrae gap for NGC\,5272 and the faint RGB of all the clusters, for which we adopt $Y=0.275$.

NGC\,5272 is the only cluster under consideration with a significant magnitude difference between the red and blue HB stars, as seen in the NGC\,5272 CMDs in 
Figs~\ref{gaia}--\ref{ps1}. We explain this difference by the different helium enrichment in agreement with findings of \citet{dalessandro2013,valcarce2016}.
However, our CMDs in Figs~\ref{gaia}--\ref{ps1} show that such an enrichment has little, if any, impact on the colour of the blue HB stars and, hence, cannot explain
any noticeable HB morphology difference.
In particular, it cannot explain the famous HB morphology difference between NGC\,6205 (M13) with its long HB blue tail (see \citetalias{ngc6205}) 
and NGC\,5272 without such a tail (see \citealt{catelan2009}).

Since this is the sixth paper in this series, many details of our analysis can be found in our previous papers. 
We refer the reader to those papers, especially to the last ones, since we believe that our studies become more refined from paper to paper.

This paper is organized as follows.
In Sect.~\ref{datasets}, we present the data sets used.
The results of our isochrone fitting are given and discussed in Sect.~\ref{results}.
We summarize our main findings and conclusions in Sect.~\ref{conclusions}.

\section{Data Sets}
\label{datasets}

For all the clusters, there are cognate data sets, which are obtained with the same telescope and/or processed within the same pipeline:
\begin{enumerate}
\item \label{filterhst} the {\it HST} Wide Field Camera 3 (WFC3) UV Legacy Survey of Galactic Globular Clusters (the $F275W$, $F336W$, and $F438W$ filters) and 
the Wide Field Channel of the Advanced Camera for Surveys (ACS; the $F606W$ and $F814W$ filters) survey of Galactic globular clusters \citep{piotto2015}, 
\citep[][hereafter NLP18]{nardiello2018} 
\footnote{\url{http://groups.dfa.unipd.it/ESPG/treasury.php}} 
(74\,205, 10\,685, 71\,049, 15\,139, and 34\,662 cluster members are used 
in NGC\,5024, NGC\,5053, NGC\,5272, NGC\,5466, and NGC\,7099, respectively) with additional photometry in the ACS $F435W$ or $F555W$ filters for 
52\,542 members of NGC\,5272 \citep{libralato2022},\footnote{\url{https://archive.stsci.edu/hlsp/hacks}}
\item \label{filtergaia} {\it Gaia} DR3 photometry in the $G$, $G_\mathrm{BP}$ and $G_\mathrm{RP}$ filters \citep{riello2021}: 
2812, 1124, 9281, 1863, and 3307 cluster members are used for 
NGC\,5024, NGC\,5053, NGC\,5272, NGC\,5466, and NGC\,7099, respectively,\footnote{The DSED isochrones for {\it Gaia} DR2 are equally suitable for DR3.}
\item \label{filterstetson} $UBVRI$ photometry from various ground-based telescopes processed by 
\citet[][hereafter SPZ19]{stetson2019}:\footnote{\url{http://cdsarc.u-strasbg.fr/viz-bin/cat/J/MNRAS/485/3042} 
with recent updates at \url{https://www.canfar.net/storage/vault/list/STETSON/homogeneous/Latest_photometry_for_targets_with_at_least_BVI}} 
4467, 1458, 10\,192, 2156, and 3186 cluster members, common in \citetalias{stetson2019} and {\it Gaia} DR3 (hereafter {\it Gaia}-induced members), are used for 
NGC\,5024, NGC\,5053, NGC\,5272, NGC\,5466, and NGC\,7099, respectively, or 9713 and 15\,078 stars of the original \citetalias{stetson2019} data sets are used for 
NGC\,5053 and NGC\,5466, respectively,\footnote{We use the \citetalias{stetson2019} and other original data sets for NGC\,5053 and NGC\,5466 as an alternative of 
{\it Gaia}-induced members from these data sets, since these distant and loose clusters have not so many {\it Gaia} DR3 stars, whereas, on the other hand, these clusters have 
few contaminants due to their high latitude (see NGC\,5053 CMD in Fig.~\ref{stetson}) and, hence, the original data sets, without any member selection, provides a reliable 
isochrone fitting with the same results as for their {\it Gaia}-induced members.} 
\item \label{filterps1} Panoramic Survey Telescope and Rapid Response System Data Release I (Pan-STARRS, PS1) photometry in the 
$g_\mathrm{PS1}$, $r_\mathrm{PS1}$, $i_\mathrm{PS1}$, $z_\mathrm{PS1}$, and $y_\mathrm{PS1}$ filters \citep{chambers2016}: 
3248, 1307, 9864, 2174, and 3426 {\it Gaia}-induced members are used for NGC\,5024, NGC\,5053, NGC\,5272, 
NGC\,5466, and NGC\,7099, respectively,\footnote{NGC\,7099 is not far from the declination limit of PS1 at about $-30\degr$ and, hence, it demonstrates less accurate 
PS1 photometry.}
\end{enumerate}

The following data sets exist for some but not all the clusters:
\begin{enumerate}
\setcounter{enumi}{4}
\item \label{filtersdss} Sloan Digital Sky Survey (SDSS) photometry in the $u_\mathrm{SDSS}$, $g_\mathrm{SDSS}$, $r_\mathrm{SDSS}$, $i_\mathrm{SDSS}$, and $z_\mathrm{SDSS}$ filters 
\citep{an2008}:\footnote{\url{http://classic.sdss.org/dr6/products/value_added/anjohnson08_clusterphotometry.htm.} We correct the SDSS magnitudes following \citet{an2009} and 
\citet{eisenstein2006}.}
3365, 1308, 9336, and 3545 {\it Gaia}-induced members of NGC\,5024, NGC\,5053, NGC\,5272, and NGC\,5466, respectively, or 
4162 and 8530 stars of the original SDSS data set for NGC\,5053 and NGC\,5466, respectively,
\item \label{filtersmss} SkyMapper Southern Sky Survey DR3 (SMSS, SMSS DR3) photometry in the $g_\mathrm{SMSS}$, $r_\mathrm{SMSS}$, $i_\mathrm{SMSS}$, and $z_\mathrm{SMSS}$ filters 
\citep{onken2019},\footnote{\url{https://skymapper.anu.edu.au}} for 1729 {\it Gaia}-induced members of NGC\,7099,
\item \label{filterwfpc2} Photometry in the $F439W$ and $F555W$ filters from the {\it HST} Wide Field and Planetary Camera 2 (WFPC2) 
\citep{piotto2002}\footnote{\url{http://groups.dfa.unipd.it/ESPG/hstphot.html}} for 9526 and 4698 stars in NGC\,5024 and NGC\,7099, respectively,
\item \label{filterwise} {\it Wide-field Infrared Survey Explorer (WISE}; \citealt{wise}) photometry in the $W1$ filter from the unWISE catalogue 
\citep{unwise}\footnote{\url{https://cdsarc.cds.unistra.fr/viz-bin/cat/II/363}}
for 902 and 444 {\it Gaia}-induced members of NGC\,5272 and NGC\,7099, respectively 
(the unWISE photometry for the remaining, more distant clusters is not deep enough for reliable results),
\item \label{filterrey1998} $BV$ photometry of 
23\,431 stars of the original data set of \citet{rey1998}\footnote{\url{https://cdsarc.cds.unistra.fr/viz-bin/cat/J/AJ/116/1775}} or its 616 {\it Gaia}-induced members of 
NGC\,5024 obtained with the University of Hawaii 2.2-m telescope, Mauna Kea,
\item \label{filtersarajedini1995} $VI$ photometry of 539 {\it Gaia}-induced members of NGC\,5053 obtained with the 0.9-m telescope at Kitt Peak National Observatory (KPNO) 
and the 1.2-m telescope at Whipple Observatory, Mt. Hopkins, Arizona \citep{sarajedini1995},\footnote{\url{https://cdsarc.cds.unistra.fr/viz-bin/cat/J/AJ/109/269}; 
$B$ photometry is not deep enough.}
\item \label{filtergrundahl1999} Str\"omgren $uvby$ photometry of 16\,403 stars of the original data set of \citet[][hereafter GCL99]{grundahl1999} or 
its 666 {\it Gaia}-induced members of NGC\,5272 obtained with the Nordic Optical Telescope (NOT), La Palma,
\item \label{filtermassari2016} Str\"omgren $uvby$ photometry of 4703 {\it Gaia}-induced members of NGC\,5272 obtained with the Isaac Newton Telescope Wide Field Camera (INT-WFC), 
La Palma \citep{massari2016},\footnote{\url{https://cdsarc.cds.unistra.fr/viz-bin/cat/J/MNRAS/458/4162}}
\item \label{filterbuonanno1994} $BV$ photometry of 
10\,540 stars of the original data set of \citet{sandage1953} and \citet{buonanno1994}\footnote{\url{https://cdsarc.cds.unistra.fr/viz-bin/cat/J/A+A/290/69}} or its 2194 
{\it Gaia}-induced members of NGC\,5272 obtained from photographic plates taken in the early 1950ths at Mt. Palomar and Mt. Wilson Observatories, 
\item \label{filterrey2001} $BV$ photometry of 2294 stars in NGC\,5272 with the 2.4-m telescope at Michigan-Dartmouth-MIT (MDM) Observatory 
\citep{rey2001},\footnote{\url{https://cdsarc.cds.unistra.fr/viz-bin/cat/J/AJ/122/3219}}
\item \label{filterbeccari2013} $BV$ photometry of
13\,955 stars of the original data set of \citet{beccari2013}\footnote{\url{https://www.lbto.org}} or its 1907 {\it Gaia}-induced members of NGC\,5466 
acquired through the blue channel of the Large Binocular Camera (LBC-blue) mounted on the Large Binocular Telescope (LBT), Mount Graham, Arizona, 
\item \label{filterfekadu2007} $BVI$ photometry of 4708 stars of the original data set of \citet{fekadu2007} or its 2020 {\it Gaia}-induced members of NGC\,5466 obtained
 with the 0.9-m telescope at KPNO,
\item \label{filterjeon2004} $BV$ photometry of 10\,633 stars of the original data set of \citet{jeon2004} or its 1783 {\it Gaia}-induced members of NGC\,5466 obtained
 with the 1.8-m telescope at the Bohyunsan Optical Astronomy Observatory in Korea,
\item \label{filtersandquist1999} $VI$ photometry of 22\,877 stars 
of the original data set of \citet{sandquist1999},\footnote{\url{https://cdsarc.cds.unistra.fr/viz-bin/cat/J/ApJ/518/262}} or its 3135 {\it Gaia}-induced members of NGC\,7099
obtained with the 4-m telescope at Cerro Tololo Inter-American Observatory (CTIO), 
\item \label{filter2mass} $J_\mathrm{2MASS}$ photometry of 945 {\it Gaia}-induced members of NGC\,7099 obtained by \citet{cohen2015} with Infrared Side Port Imager (ISPI) 
mounted on the 4-m Blanco telescope at CTIO and calibrated by use of the Two Micron All-Sky Survey (2MASS, \citealt{2mass}) stars,
\item \label{filtervista} $J_\mathrm{VISTA}$ photometry with the VISTA Hemisphere Survey with the VIRCAM instrument on the Visible and Infrared 
Survey Telescope for Astronomy (VISTA, VHS DR5) \citep{vista}\footnote{\url{https://cdsarc.cds.unistra.fr/viz-bin/ReadMe/II/367?format=html&tex=true}}
for 3300 {\it Gaia}-induced members of NGC\,7099,
\item \label{filterukidss} $J_\mathrm{UKIDSS}$ photometry with the United Kingdom Infrared Telescope Infrared Deep Sky Survey (UKIDSS) \citep{ukidss}\footnote{\url{http://www.ukidss.org};
DSED and BaSTI do not provide isochrones for the $J_\mathrm{VISTA}$ and $J_\mathrm{UKIDSS}$ filters, respectively, hence, we substitute these similar filters by each other.
We do not use other VISTA and UKIDSS filters, since they cannot be substituted.}
for 6993 {\it Gaia}-induced members of NGC\,5272.
\end{enumerate}

\begin{table}
\def\baselinestretch{1}\normalsize\normalsize
\caption[]{The Adopted Effective Wavelength $\lambda_\mathrm{eff}$ (nm) for the Filters under Consideration, Numbers of the Data Sets for which the Filters are Used,
and Photometric Uncertainty Cut (mag) Applied.
}
\label{filters}
\[
\begin{tabular}{lrlc}
\hline
\noalign{\smallskip}
Filter & $\lambda_\mathrm{eff}$ & Data sets & Cut \\
\hline
\noalign{\smallskip}
{\it HST}/WFC3 $F275W$             & 285 & \ref{filterhst}                                         & 0.08 \\
{\it HST}/WFC3 $F336W$             & 340 & \ref{filterhst}                                         & 0.07 \\
Str\"omgren $u$                    & 349 & \ref{filtergrundahl1999}, \ref{filtermassari2016}       & 0.08 \\
SDSS $u_\mathrm{SDSS}$             & 360 & \ref{filtersdss}                                        & 0.20 \\
Landolt $U$                        & 366 & \ref{filterstetson}                                     & 0.06 \\
Str\"omgren $v$                    & 414 & \ref{filtergrundahl1999}, \ref{filtermassari2016}       & 0.06 \\
{\it HST}/ACS $F435W$              & 434 & \ref{filterhst}                                         & 0.02  \\
{\it HST}/WFC3 $F438W$             & 438 & \ref{filterhst}                                         & 0.06  \\
{\it HST}/WFPC2 $F439W$            & 452 & \ref{filterwfpc2}                                       & 0.10 \\
Landolt $B$                        & 452 & \ref{filterstetson}, \ref{filterrey1998}, \ref{filterbuonanno1994}--\ref{filterjeon2004}  & 0.06 \\
Str\"omgren $b$                    & 467 & \ref{filtergrundahl1999}, \ref{filtermassari2016}       & 0.04  \\
SDSS $g_\mathrm{SDSS}$             & 471 & \ref{filtersdss}                                        & 0.06  \\  
PS1 $g_\mathrm{PS1}$               & 496 & \ref{filterps1}                                         & 0.07  \\  
{\it Gaia} DR3 $G_\mathrm{BP}$     & 505 & \ref{filtergaia}                                        & 0.10  \\
SMSS $g_\mathrm{SMSS}$             & 514 & \ref{filtersmss}                                        & 0.06  \\
{\it HST}/ACS $F555W$              & 541 & \ref{filterhst}                                         & 0.02  \\
Str\"omgren $y$                    & 548 & \ref{filtergrundahl1999}, \ref{filtermassari2016}       & 0.04  \\
{\it HST}/WFPC2 $F555W$            & 551 & \ref{filterwfpc2}                                       & 0.10  \\
Landolt $V$                        & 552 & \ref{filterstetson}, \ref{filterrey1998}, \ref{filtersarajedini1995}, \ref{filterbuonanno1994}--\ref{filtersandquist1999}  & 0.06 \\
{\it HST}/ACS $F606W$              & 599 & \ref{filterhst}                                         & 0.05  \\
{\it Gaia} DR3 $G$                 & 604 & \ref{filtergaia}                                        & 0.02  \\
SMSS $r_\mathrm{SMSS}$             & 615 & \ref{filtersmss}                                        & 0.06  \\
SDSS $r_\mathrm{SDSS}$             & 621 & \ref{filtersdss}                                        & 0.06  \\
PS1 $r_\mathrm{PS1}$               & 621 & \ref{filterps1}                                         & 0.06  \\
Landolt $R$                        & 659 & \ref{filterstetson}                                     & 0.15     \\
SDSS $i_\mathrm{SDSS}$             & 743 & \ref{filtersdss}                                        & 0.08  \\
PS1 $i_\mathrm{PS1}$               & 752 & \ref{filterps1}                                         & 0.05  \\
{\it Gaia} DR3 $G_\mathrm{RP}$     & 770 & \ref{filtergaia}                                        & 0.10  \\
SMSS $i_\mathrm{SMSS}$             & 776 & \ref{filtersmss}                                        & 0.06  \\
{\it HST}/ACS $F814W$              & 807 & \ref{filterhst}                                         & 0.05  \\
Landolt $I$                        & 807 & \ref{filterstetson}, \ref{filtersarajedini1995}, \ref{filterfekadu2007}, \ref{filtersandquist1999}  & 0.07  \\
PS1 $z_\mathrm{PS1}$               & 867  & \ref{filterps1}                                        & 0.08  \\
SDSS $z_\mathrm{SDSS}$             & 885  & \ref{filtersdss}                                       & 0.20 \\
SMSS $z_\mathrm{SMSS}$             & 913  & \ref{filtersmss}                                       & 0.06 \\
PS1 $y_\mathrm{PS1}$               & 971  & \ref{filterps1}                                        & 0.14  \\
2MASS $J_\mathrm{2MASS}$           & 1234 & \ref{filter2mass}                                      & 0.08  \\
UKIDSS $J_\mathrm{UKIDSS}$         & 1250 & \ref{filterukidss}                                     & 0.12  \\
VISTA $J_\mathrm{VISTA}$           & 1277 & \ref{filtervista}                                      & 0.10  \\
{\it WISE} $W1$                    & 3317 & \ref{filterwise}                                       & 0.15  \\
\hline
\end{tabular}
\]
\end{table}

For some clusters, we use more {\it Gaia}-induced members for the \citetalias{stetson2019}, PS1, SDSS, VISTA, and UKIDSS CMDs than {\it Gaia} cluster members for the {\it Gaia} CMDs, 
since cluster members with less precise {\it Gaia} photometry are used in the former but not in the latter case.

All the data sets with the same filters are independent, e.g. \citetalias{grundahl1999} and \citet{massari2016}.
The \citetalias{stetson2019} data sets contain photometry from various initial data sets, but not from the others under consideration.

Note that some data sources used in our previous studies provide no enough reliable data for the clusters under consideration: e.g.
Parallel-Field Catalogues of the {\it HST} UV Legacy Survey of Galactic Globular Clusters \citep{simioni2018}.

In total, 25, 25, 33, 25, and 27 filters are used for NGC\,5024, NGC\,5053, NGC\,5272, NGC\,5466, and NGC\,7099, respectively. 
Each star has photometry in some but not all filters.
Table~\ref{filters} presents the effective wavelength $\lambda_\mathrm{eff}$ in nm for the used filters, their correspondence with data sets, and the photometric uncertainty cut level.
We set the cut level as $3\,\sigma$ of the average photometric uncertainty $\sigma$ as stated by the authors of the data set,
considering that the distribution of photometric uncertainty across each data set is nearly Gaussian.
As an exception, we increase the cut level to 0.15 mag for the {\it WISE} $W1$ filter to ensure adequate representation of the TO and bright MS stars.
The median uncertainty in photometry across all filters, derived from the data set authors' uncertainty statements, is a few hundredths of a magnitude.
Generally, UV and IR photometry is less precise than optical photometry.
The uncertainty statements are utilized to assess the statistical uncertainty of our results, though it is demonstrated in Sect.~\ref{results} that the systematic uncertainty is higher.

For cleaning of the data sets, we generally follow the recommendations of their authors to select single star-like objects with reliable photometry.
To clean the {\it HST} WFC3 and ACS data sets, we select stars with $|{\tt sharp}|<0.15$, membership probability $>0.9$ or $-1$, and quality fit $>0.9$.
For the \citetalias{stetson2019}, \citetalias{grundahl1999}, SDSS, and other data sets with the stated $\chi$ and {\tt sharp} parameters, we select stars with 
$\chi<3$ and $|{\tt sharp}|<0.3$.
For the SMSS DR3 data set, we select star-like objects (i.e. with ClassStar$>0.5$) and with flags $<8$.
In the {\it Gaia} data sets we leave only stars with \verb|duplicated_source|$=0$ (\verb|Dup=0|), i.e. sources without multiple source identifiers; 
\verb|astrometric_excess_noise|$<1$ ($\epsilon i<1$);
a renormalized unit weight error not exceeding $1.4$ (\verb|RUWE|$<1.4$); 
and a corrected excess factor \verb"phot_bp_rp_excess_factor" (i.e. \verb"E(BP/RP)Corr") between $-0.14$ and $0.14$ \citep{riello2021}.

\begin{figure*}
\includegraphics{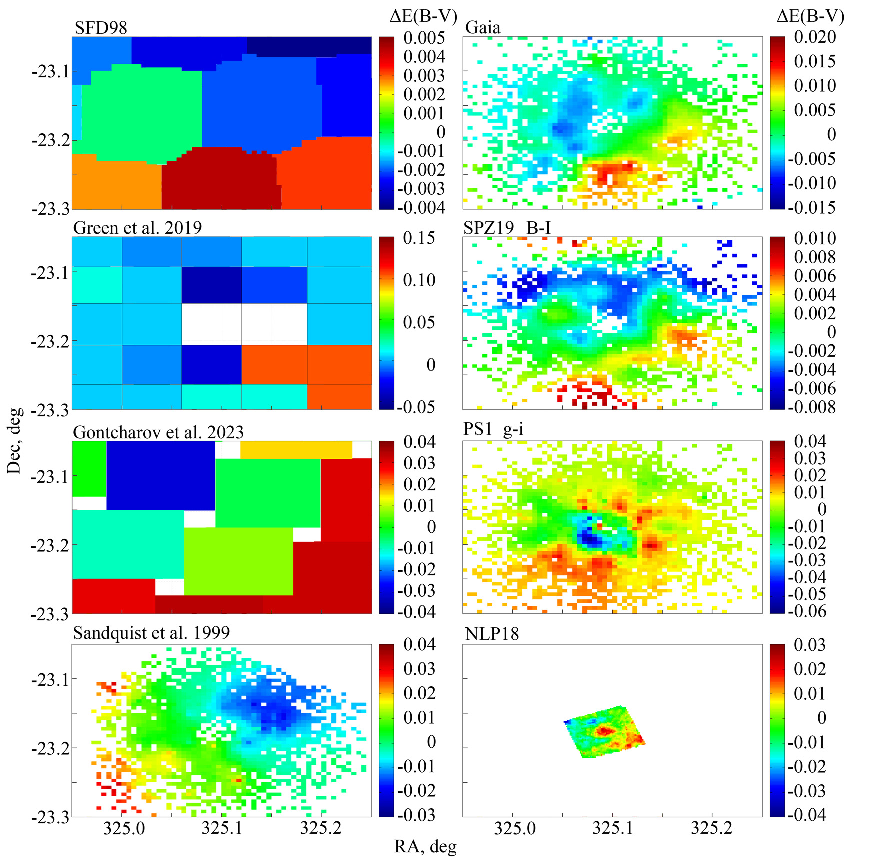}
\caption{The DR maps, converted into $\Delta E(B-V)$ using the \citetalias{ccm89} extinction law with $R_\mathrm{V}=3.1$, for the same NGC\,7099 field 
from the reddening maps of \citetalias{sfd98}, \citet{green2019}, and \citet{gmk2023}
and the data sets of {\it Gaia}, \citetalias{stetson2019}, PS1, \citet{sandquist1999}, and \citetalias{nardiello2018}
The white areas have no estimates.
}
\label{ngc7099dr}
\end{figure*}

Some systematic errors dominate in the final uncertainties. This is evident from our comparison of cross-identified data set with the same or similar filters.
For example, the \citetalias{stetson2019} and other $BVI$ data sets typically show systematic differences up to 0.04\,mag in their colours and magnitudes.
Such differences are common, expected, and well-known (see \citealt{fekadu2007}, \citetalias{stetson2019} and our previous papers).
Some of the differences appear as systematic variations of CMD over cluster field probably due to photometry zero-point variations, point-spread function variations, 
telescope focus change, distortion, telescope breathing, stellar generation variations, and other field systematics, discussed by \citet{anderson2008}.
Consequently, these systematics are difficult to separate from DR.
Accordingly, the rather high DR in the fields of these clusters (see Table~\ref{properties}) should be considered as a manifestation of the systematics.
Hence, the higher the DR from a data set, the higher its systematics. Therefore, DR investigation can be used to briefly estimate a systematic quality of the data sets.
However, it can be applied only to data sets with sufficient coverage of the cluster fields, i.e. at least 2000 stars.
Similar to \citetalias{ngc6397}, we estimate DR following the method of \citetalias{bonatto2013}.
Our DR corrections are generally within $\pm0.06$\,mag. 

It is seen, for example, in Fig.~\ref{ngc7099dr} with some DR maps for the NGC\,7099 field derived from global reddening maps and data sets used.
The DR maps do not agree in the DR value (due to systematics in the data) and, hence, we have to show them in Fig.~\ref{ngc7099dr} on different scales.
However, a qualitative similarity is seen for all maps, except that from the \citet{sandquist1999} data set, which is a reason not to use the latter for our final results.
Generally, reddening increases from the upper left (North-West) to the lower right (South-East) corner. 
This is seen even in the \citetalias{nardiello2018} map covering only the centre of NGC\,7099.
This can be partially due to an influence of a bright star 41~Cap of $V\approx5.2$\,mag located about 23 arcmin southeast of NGC\,7099, 
i.e. outside the field shown in Fig.~\ref{ngc7099dr}, but not far from its lower right corner.
It is worth noting about the map estimates that a very high reddening peak in the lower right corner of the \citet{green2019} map may be due to some feature of their method in GC 
fields (Green, private communication),
whereas \citetalias{sfd98} probably underestimates the reddening gradient in the cluster field due to a significant gradient of dust temperature in this field.

Among the clusters under consideration, NGC\,7099 demonstrates the highest reddening, albeit only slightly higher than the systematics in the data sets.
That is why we can see the real NGC\,7099 DR manifest against a background of the systematics.
In contrast, the latter completely dominates in the DR maps of the remaining clusters and, hence, these maps have little to do with each other as for different data sets, 
as for different CMDs/colours of the same data set.


Correction for DR reduces the scatter of stars around their ridge lines or best-fitting isochrones in CMDs.
Note that the mean DR correction for each CMD is exactly zero. 
A smooth and rather low DR for the {\it HST}, {\it Gaia}, \citetalias{stetson2019}, PS1, SDSS, SMSS, \citetalias{grundahl1999}, and \citet{massari2016} data sets makes them 
the key data sets for deriving cluster parameters.
In contrast, some other data sets demonstrate much more noticeable systematics: e.g. we have to eliminate the brightest stars from the data sets of \citet{buonanno1994} ($V<15.2$), 
\citet{beccari2013} ($V<16$), and VISTA ($J_\mathrm{VISTA}<12$) due to their unacceptable systematics.

We fit isochrones to a hundred CMDs with different colours. As in our previous papers, the results for adjacent CMDs appear consistent,
and they are more reliable for CMDs in the optical range (i.e. with filters within $430<\lambda_\mathrm{eff}<1000$ nm) than for UV, UV--optical, optical--IR, and IR CMDs, 
such as \citetalias{grundahl1999} $u-v$, \citetalias{stetson2019} $U-B$, {\it Gaia}--VISTA, \citetalias{stetson2019}--UKIDSS, PS1--unWISE, UKIDSS--unWISE, and others.
Furthermore, uncertainties of the derived reddenings are dominated by the data set systematics, which are irrespective to a wavelength range under consideration in the optical range, 
while increase in the UV or IR.
Therefore, for these low-reddening clusters, 
the most reliable reddening estimates with the lowest relative uncertainties can be derived from the widest optical wavelength range CMD of each data set.
Thus, we derive the final cluster parameters by use of the only key optical CMD for each key data set:
(i) $F606W-F814W$ from \citetalias{nardiello2018}, (ii) $G_\mathrm{BP}-G_\mathrm{RP}$ from {\it Gaia} DR3, (iii) $B-I$ from \citetalias{stetson2019}, 
(iv) $g_\mathrm{PS1}-i_\mathrm{PS1}$ from PS1 for all the clusters, and additionally
(v) $g_\mathrm{SDSS}-i_\mathrm{SDSS}$ from SDSS for all clusters, except NGC\,7099, or (v) $g_\mathrm{SMSS}-i_\mathrm{SMSS}$ from SMSS for NGC\,7099, and also 3 data sets for NGC\,5272:
(vi) $F435W-F814W$ from \citet{libralato2022}, (vii) $b-y$ from \citetalias{grundahl1999}, and (viii) $b-y$ from \citet{massari2016}.
Thus, we have 5 key CMDs for all clusters, except NGC\,5272, for which we have 8 key CMDs.

The remaining data sets and CMDs are also important, since we use their estimates of cluster parameters to evaluate the systematic uncertainty of our results.
For example, the key CMDs provide [Fe/H]=-1.95, age 12.15 Gyr, $R=15.59$ kpc, and $E(B-V)=0.023$ for NGC\,5466, while
the addition of the CMDs of \citet{jeon2004}, \citet{fekadu2007}, and \citet{beccari2013} provides [Fe/H]=-1.92, age 12.03 Gyr, $R=15.75$ kpc, and $E(B-V)=0.017$.
The differences of these estimates determine the lower limit of their systematic uncertainties: 0.03 dex, 0.12 Gyr, 0.16 kpc, and 0.006\,mag, respectively.

As in our previous studies, we also use the cross-identification of data sets to convert the derived reddenings into extinction for each filter we consider, 
and draw an empirical extinction law (i.e. a dependence of extinction on wavelength) for each combination of cluster, data set, and model.
However, for such low-extinction/reddening clusters, empirical extinction law, being based on a ratio of a low extinction to a low reddening, is very uncertain and it is strongly 
affected by systematic errors of the data sets. These errors appear as deviations of the extinction estimates from an average law in agreement with the systematic 
colour differences in the direct comparison of the data sets.
Therefore, we draw the empirical extinction laws using all independent CMDs for the whole wavelength range from the UV to IR 
in order to verify only whether they agree within the extinction uncertainties with a common \citet[][hereafter CCM89]{ccm89} extinction law 
with $R_\mathrm{V}=3.1$.\footnote{Extinction-to-reddening ratio $R_\mathrm{V}\equiv A_\mathrm{V}/E(B-V)=3.1$ is defined for early type MS stars, 
while the observed ratio $A_\mathrm{V}/E(B-V)$ depends on intrinsic spectral energy distribution of stars under consideration \citep{casagrande2014}.
For rather cool and metal-poor stars of the clusters under consideration the observed extinction in the $V$ filter is calculated as $A_\mathrm{V}=3.48E(B-V)$,
while the extinction coefficients are calculated for the median effective temperature 6400 K of the cluster members.}

NGC\,5272 and NGC\,7099 have optical--IR CMDs covering a wide wavelength range and related to well-defined very low IR extinctions in the VISTA, UKIDSS or unWISE bands.
Hence, these CMDs appear especially fruitful for the testing of extinction law.
Similar to our previous papers, we calculate extinctions in optical filters from the derived reddenings and IR extinctions. For example,
\begin{equation}
\label{avaw1}
A_\mathrm{V}=(A_\mathrm{V}-A_\mathrm{W1})+A_\mathrm{W1}=E(V-W1)+A_\mathrm{W1},
\end{equation}
where $E(V-W1)$ is obtained from a CMD, while very low extinction $A_\mathrm{W1}$ in the $W1$ filter is calculated using the 
\citetalias{ccm89} extinction law with $R_\mathrm{V}=3.1$ and optical extinctions and slightly upgraded iteratively with upgrade of the optical extinctions.
Finally, the \citetalias{ccm89} extinction law with $R_\mathrm{V}=3.1$ appears acceptable for all the clusters, data sets and models.
In particular, we use this law to convert the final derived reddenings to the final extinctions $A_\mathrm{V}$.

\subsection{{\it Gaia} DR3 Cluster Members}
\label{edr3}

Table~\ref{properties} shows rather different tidal radius estimates for each cluster.
Therefore, we consider initial {\it Gaia} DR3 samples within initial radii which exceed any previous estimate.
We find empirical truncation radii, presented in Table~\ref{properties}, as the radii where the cluster star count surface density drops to the Galactic background.
All the data sets are truncated at these radii to reduce contamination from non-members.
The truncation allows us to create very clean samples, albeit incomplete.
Since all these clusters, except NGC\,7099, have tidal tails \citep{chun2010,yang2023}, we might lose several cluster members outside our truncation radii, 
but it does not affect our results.

As described in our previous studies, accurate {\it Gaia} DR3 parallaxes and PMs are used to select cluster members and derive systemic parallaxes and PMs.

As in our previous studies, the final empirical standard deviations of cluster member PMs $\sigma_{\mu_{\alpha}\cos(\delta)}$ and $\sigma_{\mu_{\delta}}$ 
are reasonable, but higher by about 20--70\% than the mean stated PM uncertainties, which may mean an underestimation of real errors in the PMs.

Our final weighted mean systemic PMs are presented in Table~\ref{systemic} in comparison to those from \citet[][hereafter VB21]{vasiliev2021} and \citet{vitral2021} 
also obtained from {\it Gaia} DR3 but by different approaches. 
Note the different nature of the stated uncertainties: statistic ones for ours and \citet{vitral2021}'s estimates, while the total (statistic plus systematic) ones for the 
\citetalias{vasiliev2021} estimates. The latter must be adopted as the final, more realistic uncertainties of our PMs.
Our PM estimates agree with those from \citetalias{vasiliev2021} within $\pm0.017$ mas\,yr$^{-1}$ and, hence, within the total PM uncertainties.
The \citet{vitral2021}'s estimates for NGC\,5024 deviate from the rest significantly, up to 0.095 and 0.062 mas\,yr$^{-1}$ for the PM components 
$\mu_{\alpha}\cos(\delta)$ and $\mu_{\delta}$, respectively.

We correct our median parallaxes of cluster members for the parallax zero-point following \citet{lindegren2021} and present them in Table~\ref{parallax} for comparison with other 
estimates in Sect.~\ref{results}.
We adopt the total uncertainty of {\it Gaia} DR3 parallaxes, found by \citetalias{vasiliev2021} as 0.01 mas.

Note that \citetalias{nardiello2018} and \citet{libralato2022} have cleaned their data sets from non-members by use of dedicated {\it HST} PMs.
For such high-latitude clusters this cleaning is almost perfect.

\begin{table}
\def\baselinestretch{1}\normalsize\normalsize
\caption[]{The Cluster Systemic PMs (mas\,yr$^{-1}$).
}
\label{systemic}
\[
\begin{tabular}{llcc}
\hline
\noalign{\smallskip}
Cluster & Source & $\mu_{\alpha}\cos(\delta)$ & $\mu_{\delta}$ \\
\hline
\noalign{\smallskip}
          & This study                & $-0.150\pm0.013$ & $-1.336\pm0.011$ \\
NGC\,5024 & \citetalias{vasiliev2021} & $-0.134\pm0.024$ & $-1.331\pm0.024$ \\
          & \citet{vitral2021}        & $-0.229\pm0.008$ & $-1.274\pm0.008$ \\
\noalign{\smallskip}
          & This study                & $-0.333\pm0.016$ & $-1.224\pm0.017$ \\
NGC\,5053 & \citetalias{vasiliev2021} & $-0.329\pm0.026$ & $-1.214\pm0.025$ \\
          & \citet{vitral2021}        & $-0.325\pm0.011$ & $-1.209\pm0.011$ \\
\noalign{\smallskip}
          & This study                & $-0.150\pm0.005$ & $-2.671\pm0.004$ \\
NGC\,5272 & \citetalias{vasiliev2021} & $-0.152\pm0.022$ & $-2.670\pm0.023$ \\
          & \citet{vitral2021}        & $-0.151\pm0.004$ & $-2.667\pm0.004$ \\
\noalign{\smallskip}
          & This study                & $-5.360\pm0.012$ & $-0.828\pm0.010$ \\
NGC\,5466 & \citetalias{vasiliev2021} & $-5.343\pm0.024$ & $-0.823\pm0.024$ \\
          & \citet{vitral2021}        & $-5.368\pm0.008$ & $-0.838\pm0.008$ \\
\noalign{\smallskip}
          & This study                & $-0.735\pm0.011$ & $-7.284\pm0.009$ \\
NGC\,7099 & \citetalias{vasiliev2021} & $-0.738\pm0.025$ & $-7.299\pm0.025$ \\
          & \citet{vitral2021}        & $-0.737\pm0.006$ & $-7.293\pm0.006$ \\
\hline
\end{tabular}
\]
\end{table}

\begin{table*}
\def\baselinestretch{1}\normalsize\small
\caption[]{Parallax Estimates (mas) with their Total (Statistic and Systematic) Uncertainties for Clusters under Consideration.
}
\label{parallax}
\[
\begin{tabular}{lccccc}
\hline
\noalign{\smallskip}
 Parallax                                             &  NGC\,5024      &  NGC\,5053      & NGC\,5272       & NGC\,5466       & NGC\,7099 \\
\hline
\noalign{\smallskip}
\citetalias{vasiliev2021}, {\it Gaia} DR3 astrometry  & $0.064\pm0.011$ & $0.047\pm0.011$ & $0.106\pm0.010$ & $0.053\pm0.011$ & $0.132\pm0.011$ \\
\citetalias{baumgardt2021}, various methods           & $0.054\pm0.001$ & $0.057\pm0.001$ & $0.098\pm0.001$ & $0.062\pm0.001$ & $0.118\pm0.001$ \\ 
This study, {\it Gaia} DR3 astrometry                 & $0.066\pm0.011$ & $0.041\pm0.015$ & $0.110\pm0.010$ & $0.062\pm0.011$ & $0.119\pm0.011$ \\
This study, isochrone fitting                         & $0.055\pm0.002$ & $0.059\pm0.002$ & $0.099\pm0.003$ & $0.064\pm0.002$ & $0.121\pm0.004$ \\ 
\hline
\end{tabular}
\]
\end{table*}

\begin{table*}
\def\baselinestretch{1}\normalsize\normalsize
\caption{The Results of our Isochrone Fitting for Two Models and Some Key CMDs. \\
In all the CMDs, the colour is the abscissa, the reddening is the colour excess, and the magnitude in the redder filter is the ordinate.
Each derived reddening is followed by 
%
corresponding $E(B-V)$, given in parentheses 
and calculated using extinction coefficients from \citet{casagrande2014,casagrande2018a,casagrande2018b} or \citetalias{ccm89} with $R_\mathrm{V}=3.1$.
[Fe/H] is given only for CMDs, which allow its calculation as a fitting parameter. 
Age is in Gyr, $R$ is in kpc.
The complete table is available online.
}
\label{cmds}
\[
\begin{tabular}{lcccccccc}                                            
\hline
\noalign{\smallskip}
   & \multicolumn{4}{c}{BaSTI} & \multicolumn{4}{c}{DSED} \\
Data set and colour                       & [Fe/H] & Age  & $R$  & Reddening   & [Fe/H] & Age  & $R$  & Reddening   \\
\hline
\noalign{\smallskip}
   & \multicolumn{8}{c}{NGC\,5024} \\
\noalign{\smallskip}
\citetalias{nardiello2018} $F606W-F814W$  & $-1.9$ & 12.5      & 18.3      & $0.018$ [0.018]      & $-1.9$ & 13.0      & 18.0      & $0.027$ [0.028] \\
{\it Gaia} $G_\mathrm{BP}-G_\mathrm{RP}$  & $-1.9$ & 13.0      & 18.0      & $0.043$ [0.031]      & $-1.9$ & 12.5      & 18.3      & $0.071$ [0.051]  \\
\citetalias{stetson2019} $B-I$            & $-1.9$ & 13.0      & 18.5      & $0.015$ [0.007]      & $-1.9$ & 13.0      & 18.3      & $0.036$ [0.017] \\
PS1 $g_\mathrm{PS1}-i_\mathrm{PS1}$       & $-2.0$ & 13.0      & 18.2      & $0.040$ [0.025]      & $-2.0$ & 13.5      & 18.0      & $0.032$ [0.022] \\
\ldots & \ldots & \ldots & \ldots & \ldots  & \ldots  & \ldots  & \ldots  & \ldots \\
\noalign{\smallskip}
   & \multicolumn{8}{c}{NGC\,5053} \\
\citetalias{nardiello2018} $F606W-F814W$  & $-2.2$ & 12.5      & 17.2      & $0.018$ [0.018]      & $-2.2$ & 12.5      & 17.2      & $0.022$ [0.023] \\
{\it Gaia} $G_\mathrm{BP}-G_\mathrm{RP}$  & $-2.1$ & 12.5      & 17.0      & $0.038$ [0.027]      & $-2.0$ & 12.5      & 17.0      & $0.055$ [0.039]  \\
\citetalias{stetson2019} $B-I$            & $-2.2$ & 12.5      & 17.2      & $0.033$ [0.017]      & $-2.1$ & 13.0      & 16.8      & $0.032$ [0.016] \\
PS1 $g_\mathrm{PS1}-i_\mathrm{PS1}$       & $-2.1$ & 13.0      & 17.0      & $0.020$ [0.013]      & $-2.0$ & 13.5      & 16.7      & $0.008$ [0.005] \\
\ldots & \ldots & \ldots & \ldots & \ldots  & \ldots  & \ldots  & \ldots  & \ldots \\
\noalign{\smallskip}
   & \multicolumn{8}{c}{NGC\,5272} \\
\citetalias{nardiello2018} $F606W-F814W$  & $-1.5$ & 11.5      & 10.1      & $0.014$ [0.014]      & $-1.5$ & 11.5      & 10.0      & $0.027$ [0.028] \\
{\it Gaia} $G_\mathrm{BP}-G_\mathrm{RP}$  & $-1.5$ & 11.5      & 10.0      & $0.031$ [0.022]      & $-1.6$ & 11.0      & 10.2      & $0.073$ [0.052]  \\
\citetalias{stetson2019} $B-I$            & $-1.6$ & 11.5      & 10.3      & $0.023$ [0.011]      & $-1.6$ & 11.5      & 10.1      & $0.061$ [0.029] \\
PS1 $g_\mathrm{PS1}-i_\mathrm{PS1}$       & $-1.7$ & 12.0      & 10.2      & $0.031$ [0.020]      & $-1.6$ & 12.0      & 10.1      & $0.031$ [0.020] \\
\ldots & \ldots & \ldots & \ldots & \ldots  & \ldots  & \ldots  & \ldots  & \ldots \\
\noalign{\smallskip}
   & \multicolumn{8}{c}{NGC\,5466} \\
\citetalias{nardiello2018} $F606W-F814W$  & $-1.9$ & 12.0      & 15.6      & $0.015$ [0.015]      & $-2.0$ & 12.5      & 15.6      & $0.024$ [0.025] \\
{\it Gaia} $G_\mathrm{BP}-G_\mathrm{RP}$  & $-1.9$ & 11.5      & 15.6      & $0.042$ [0.030]      & $-1.9$ & 12.0      & 15.6      & $0.069$ [0.049]  \\
\citetalias{stetson2019} $B-I$            & $-2.0$ & 12.5      & 15.6      & $0.038$ [0.018]      & $-2.0$ & 12.5      & 15.5      & $0.057$ [0.027] \\
PS1 $g_\mathrm{PS1}-i_\mathrm{PS1}$       & $-2.0$ & 12.0      & 15.7      & $0.029$ [0.019]      & $-1.9$ & 12.5      & 15.4      & $0.022$ [0.014] \\
\ldots & \ldots & \ldots & \ldots & \ldots  & \ldots  & \ldots  & \ldots  & \ldots \\
\noalign{\smallskip}
   & \multicolumn{8}{c}{NGC\,7099} \\
\citetalias{nardiello2018} $F606W-F814W$  & $-2.2$ & 13.0      &  8.3      & $0.040$ [0.041]      & $-2.2$ & 13.5      &  8.2      & $0.050$ [0.050] \\
{\it Gaia} $G_\mathrm{BP}-G_\mathrm{RP}$  & $-1.9$ & 12.5      &  8.2      & $0.062$ [0.044]      & $-1.9$ & 12.0      &  8.3      & $0.087$ [0.062]  \\
\citetalias{stetson2019} $B-I$            & $-2.1$ & 13.0      &  8.4      & $0.079$ [0.037]      & $-2.2$ & 13.5      &  8.3      & $0.100$ [0.047] \\
PS1 $g_\mathrm{PS1}-i_\mathrm{PS1}$       & $-2.1$ & 13.0      &  8.3      & $0.066$ [0.043]      & $-2.1$ & 13.0      &  8.4      & $0.058$ [0.038] \\
\ldots & \ldots & \ldots & \ldots & \ldots  & \ldots  & \ldots  & \ldots  & \ldots \\
\hline
\end{tabular}
\]
\end{table*}

\begin{figure*}
\includegraphics{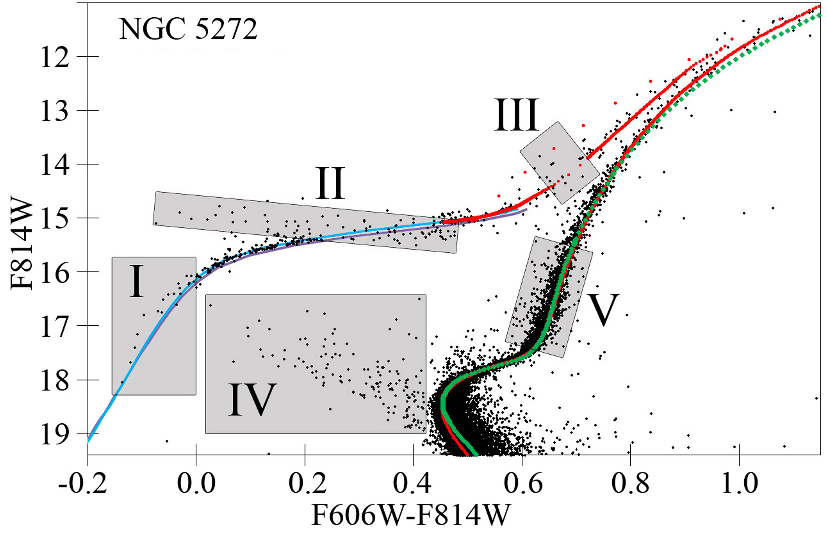}
\caption{A central part of the $F606W-F814W$ versus $F814W$ CMD from the \citetalias{nardiello2018} data set for NGC\,5272.
The isochrones from BaSTI (red) and DSED (green) for $Y=0.25$, the BaSTI ZAHB for $Y=0.25$ (purple) and for $Y=0.275$ (blue) are calculated with the best-fitting parameters 
from Table~\ref{cmds}.
The grey areas are the CMD domains of the (I) extremely blue HB, (II) RR~Lyrae, (III) other variables, (IV) blue stragglers, and (V) faint RGB, which are discussed in the text.
}
\label{hst_details}
\end{figure*}

\subsection{Isochrone-to-data Fitting}
\label{fitting}

Owing to a minor contamination of these high-latitude clusters and the accurate selection of the cluster members, the distribution of stars in our CMDs is well defined.
Therefore, as in \citetalias{ngc6397}, we fit isochrones directly to a bulk of cluster members, without calculation of a fiducial sequence.

We assign a weight to each data point to balance the contributions of different CMD domains.
The weight is inversely proportional to the number of stars of a given magnitude, i.e. it reflects the luminosity function of a given data set.
For each CMD, modern computers allow us to consider a hundred thousand sets of parameters ($Y$, [Fe/H], distance, reddening, and age) for a reasonable grid 
in their 5-dimensional space.
The grid has steps of 0.025 points, 0.1 dex, 0.05 kpc, 0.001 mag, and 0.5 Gyr for $Y$, [Fe/H], distance, reddening, and age, respectively.
For each set of parameters we calculate the sum of the squares of the residuals between the isochrones and the data points.
The best solution, presented in Table~\ref{cmds}, is the one with the minimal sum of the squares of the residuals.

As in \citetalias{ngc6397}, we have to exclude four CMD domains from the direct fitting: the extremely blue HB (i.e. the area bluer than the turn of the observed HB downward), 
RR~Lyrae, other variables, and blue stragglers, marked I, II, III, and IV in Fig.~\ref{hst_details}, respectively.
These clusters contain a lot of RR~Lyrae and other variables \citep{arellano2022,arellano2024}. 
Therefore, their detection by {\it Gaia} and \citetalias{stetson2019} and subsequent removal is very fruitful for correct determination of the HB magnitude and, hence, cluster distance.
Unfortunately, as far as we know, variables are not detected in the {\it HST} data sets. Therefore, we approximately determine the areas II and III for the {\it HST} data sets 
and eliminate all stars in them before our fitting\footnote{We cannot eliminate the variables, known from the {\it Gaia} and \citetalias{stetson2019} data sets, by their 
cross-identification with the {\it HST} data sets, since these data sets cover different parts of the cluster fields and, hence, have little, if any, common variables.}, 
but retain these stars in Fig.~\ref{hst} as an instance.
The undetected RR~Lyrae variables may bias some previous distance estimates.

Fig.~\ref{hst_details} shows that after the exclusion of these domains, these clusters have enough HB and AGB stars to determine cluster parameters: 
they are the blue HB stars between the areas I and II, red HB stars between the areas II and III, and AGB stars on the red side of the area III. 
As noted earlier, the blue HB stars of NGC\,5272 are better fitted with $Y=0.275$.
Another domain better fitted with $Y=0.275$ is the faint RGB, marked V in Fig.~\ref{hst_details}.

We present some key CMDs with isochrone fits in Figs~\ref{gaia}--\ref{ps1}. 
Other CMDs are presented online or can be provided on request.

\begin{table*}
\def\baselinestretch{1}\normalsize\normalsize
\caption[]{Our [Fe/H] (dex), Age (Gyr), Distance (kpc), Distance Modulus (mag), Apparent $V$-band Distance Modulus (mag), and $E(B-V)$ (mag) Estimates Averaged from the Key CMDs. \\
The $E(B-V)$ estimates are calculated from the derived reddenings by use of extinction coefficients from \citet{casagrande2014,casagrande2018a,casagrande2018b} or 
\citetalias{ccm89} with $R_\mathrm{V}=3.1$.
The uncertainties after the values are standard deviations of one estimate.
`$\sigma$', `Model $\Delta$' and `Systematics' are the standard deviation of the average value, half the difference between the models, and systematic uncertainty of the average value,
respectively.
}
\label{estimates}
\[
\begin{tabular}{lcccccc}
\hline
\noalign{\smallskip}
Cluster    & BaSTI           & DSED          &  Average value & $\sigma$ & Model $\Delta$ & Systematic \\    
\hline
\noalign{\smallskip}
          \multicolumn{6}{c}{[Fe/H]}   \\
\noalign{\smallskip}
NGC\,5024  & $-1.92\pm0.04$ & $-1.94\pm0.05$ & $-1.93\pm0.05$ & 0.02 & 0.01 & 0.10 \\  
NGC\,5053  & $-2.12\pm0.08$ & $-2.04\pm0.11$ & $-2.08\pm0.10$ & 0.03 & 0.04 & 0.10 \\ 
NGC\,5272  & $-1.58\pm0.07$ & $-1.63\pm0.09$ & $-1.60\pm0.08$ & 0.02 & 0.03 & 0.10 \\ 
NGC\,5466  & $-1.96\pm0.05$ & $-1.94\pm0.05$ & $-1.95\pm0.05$ & 0.02 & 0.01 & 0.10 \\
NGC\,7099  & $-2.04\pm0.13$ & $-2.10\pm0.12$ & $-2.07\pm0.13$ & 0.04 & 0.03 & 0.10 \\
\noalign{\smallskip}
        \multicolumn{6}{c}{Age}    \\
\noalign{\smallskip}
NGC\,5024  & $12.90\pm0.22$ & $13.10\pm0.42$ & $13.00\pm0.33$ & 0.11 & 0.10 & 0.80 \\  
NGC\,5053  & $12.60\pm0.22$ & $12.80\pm0.45$ & $12.70\pm0.35$ & 0.11 & 0.10 & 0.80  \\ 
NGC\,5272  & $11.69\pm0.26$ & $11.56\pm0.32$ & $11.63\pm0.29$ & 0.07 & 0.06 & 0.80 \\ 
NGC\,5466  & $12.00\pm0.35$ & $12.30\pm0.27$ & $12.15\pm0.34$ & 0.11 & 0.15 & 0.80 \\
NGC\,7099  & $12.70\pm0.45$ & $12.90\pm0.65$ & $12.80\pm0.54$ & 0.17 & 0.10 & 0.80 \\
\noalign{\smallskip}
        \multicolumn{6}{c}{Distance} \\
\noalign{\smallskip}
NGC\,5024  & $18.28\pm0.19$ & $18.16\pm0.15$ & $18.22\pm0.18$ & 0.06 & 0.06 & 0.60 \\  
NGC\,5053  & $17.08\pm0.11$ & $16.90\pm0.20$ & $16.99\pm0.18$ & 0.06 & 0.09 & 0.56 \\ 
NGC\,5272  & $10.11\pm0.16$ & $10.05\pm0.16$ & $10.08\pm0.16$ & 0.04 & 0.03 & 0.33 \\
NGC\,5466  & $15.64\pm0.05$ & $15.54\pm0.09$ & $15.59\pm0.09$ & 0.03 & 0.05 & 0.51 \\
NGC\,7099  & $8.26\pm0.11$  & $8.32\pm0.08$  & $8.29\pm0.10$ & 0.03 & 0.03 & 0.27 \\
\noalign{\smallskip}
        \multicolumn{6}{c}{Distance modulus} \\
\noalign{\smallskip}
NGC\,5024  & $16.31\pm0.02$ & $16.30\pm0.02$ & $16.30\pm0.02$ & 0.01 & 0.01 & 0.07 \\  
NGC\,5053  & $16.16\pm0.01$ & $16.14\pm0.03$ & $16.15\pm0.02$ & 0.01 & 0.01 & 0.07 \\ 
NGC\,5272  & $15.02\pm0.03$ & $15.01\pm0.04$ & $15.02\pm0.03$ & 0.01 & 0.01 & 0.07 \\
NGC\,5466  & $15.97\pm0.01$ & $15.96\pm0.01$ & $15.96\pm0.01$ & 0.01 & 0.01 & 0.07 \\
NGC\,7099  & $14.58\pm0.03$ & $14.60\pm0.02$ & $14.59\pm0.03$ & 0.01 & 0.01 & 0.07 \\
\noalign{\smallskip}
        \multicolumn{6}{c}{Apparent $V$-band distance modulus} \\
\noalign{\smallskip}
NGC\,5024  & $16.37\pm0.02$ & $16.39\pm0.05$ & $16.38\pm0.04$ & 0.01 & 0.01 & 0.08 \\  
NGC\,5053  & $16.21\pm0.04$ & $16.20\pm0.07$ & $16.21\pm0.05$ & 0.02 & 0.01 & 0.08 \\ 
NGC\,5272  & $15.07\pm0.03$ & $15.13\pm0.06$ & $15.10\pm0.05$ & 0.01 & 0.03 & 0.08 \\
NGC\,5466  & $16.04\pm0.02$ & $16.05\pm0.05$ & $16.05\pm0.04$ & 0.01 & 0.01 & 0.08 \\
NGC\,7099  & $14.73\pm0.03$ & $14.77\pm0.03$ & $14.75\pm0.03$ & 0.01 & 0.02 & 0.08 \\
\noalign{\smallskip}
        \multicolumn{6}{c}{$E(B-V)$} \\
\noalign{\smallskip}
NGC\,5024  & $0.018\pm0.011$ & $0.028\pm0.013$ & $0.023\pm0.013$ & 0.004 & 0.005 & 0.010 \\
NGC\,5053  & $0.015\pm0.010$ & $0.018\pm0.014$ & $0.017\pm0.011$ & 0.004 & 0.002 & 0.010 \\
NGC\,5272  & $0.013\pm0.007$ & $0.033\pm0.016$ & $0.023\pm0.016$ & 0.004 & 0.010 & 0.010 \\
NGC\,5466  & $0.019\pm0.006$ & $0.027\pm0.013$ & $0.023\pm0.011$ & 0.003 & 0.004 & 0.010 \\
NGC\,7099  & $0.041\pm0.003$ & $0.048\pm0.009$ & $0.045\pm0.007$ & 0.002 & 0.003 & 0.010 \\
\hline
\end{tabular}
\]
\end{table*}


\section{Results}
\label{results}

Table~\ref{estimates} presents our estimates of [Fe/H], age, distance, distance modulus $(m-M)_0$, apparent $V$-band distance modulus $(m-M)_\mathrm{V}$, and $E(B-V)$ 
averaged for the key CMDs. 
We provide the uncertainties after the values as standard deviations of one estimate in order to emphasize a good agreement between the models seen in the fact that the 
standard deviation of the united BaSTI plus DSED sample is comparable with the model standard deviations.
The standard deviation of the average value (i.e. the standard deviation of one estimate divided by the square root of the number of the estimates), 
half the difference between the model estimates, and systematic uncertainty of the average value are given in separate columns.

As in \citetalias{ngc6397}, we assign the systematic uncertainty to our [Fe/H] and age estimates as $0.1$ dex and 0.8 Gyr, respectively. 
The systematic uncertainties of distance and apparent $V$-band distance modulus are calculated from that of distance modulus, which is assigned as 0.07\,mag.
This is a conservative estimate of all possible systematic effects on the magnitude of the HB and AGB stars, which is the basis for distance estimates.

The systematic uncertainty of [Fe/H] is the dominant contributor to the systematic uncertainties of our reddening and extinction estimates, resulting in values equivalent to 
$\sigma_{E(B-V)}=0.01$ and $\sigma_{A_\mathrm{V}}=0.03$ mag, respectively.
The former is presented in Table~\ref{estimates} as `Systematics'.
The systematic uncertainties of our reddening estimates, arising from imperfections in the models, can be assessed by calculating half the difference between the models presented in 
Table~\ref{estimates} as `Model $\Delta$'. 
It is seen that this model systematics is less significant than that of [Fe/H], with the exception of NGC\,5272.
It is worth noting that the absence of highly negative derived reddenings (see Table~\ref{cmds}) confirms the reliability of the isochrones, at least, for the optical filters.

Our [Fe/H] estimates support lower [Fe/H] estimates from the literature for NGC\,5272 [such as the one of \citet{arellano2024} from Table~\ref{properties}, based on their
RR~Lyrae calibration, and all from Table~\ref{diversity}, except the one of \citet{stenning2016}], 
while higher [Fe/H] estimates for the remaining low-metallicity clusters [such as those of \citet{meszaros2020}, photometric ones of \citet{jurcsik2023}, and the ones 
of \citet{arellano2024}, based on their RR~Lyrae calibration, from Table~\ref{properties}, as well as some from Table~\ref{diversity}].
Accordingly, our estimates support the arguments of \citet{mucciarelli2020} in favour of photometrically and against spectroscopically derived [Fe/H] of low-metallicity GCs.

Also, we conclude that three clusters (NGC\,5024, NGC\,5053, and NGC\,7099) have nearly the same old age, NGC\,5466 is younger, while NGC\,5272 is much younger.
This qualitatively agrees with only \citet{dotter2010} among the estimates in Table~\ref{properties}, albeit NGC\,5272 always tends to be rather young.
Age estimates from Table~\ref{properties} are good representatives of recent age estimates from the literature, which are not consistent even in relative age estimates: 
for example, NGC\,5272 is older than NGC\,5466 by \citet{valcin2020}, while sligthly younger by \citet{dotter2010}
and much younger by \citet{forbes2010}; NGC\,7099 is older than NGC\,5024 and NGC\,5053 by \citet{vandenberg2013}, while it is younger by \citet{valcin2020}.
Anyway, our age estimates agree with all the estimates within their uncertainties.

Since distance estimates from the literature (e.g. in a comprehensive compilation of \citetalias{baumgardt2021}) are rather diverse for these clusters, 
our distance estimates agree with some of them, while disagree with some others.
Examples of former and latter, presented in Table~\ref{properties}, are the estimates of \citet{arellano2024}, obtained 
from their calibrations for the RRc stars,\footnote{A good agreement of our estimates with those of \citet{arellano2024} for both $R$ and [Fe/H] suggests that 
his calibrations for RRc variables are robust, whereas his recommendation to calibrate RRab and RRc variables separately is valid.}
and \citet{hunt2023} obtained by a sophisticated method.
However, the most important is that our distance estimates agree with the most probable compiled distance estimates of \citetalias{baumgardt2021} presented in Table~\ref{properties} 
within $1.0\sigma$, $1.8\sigma$, $0.9\sigma$, $2.5\sigma$, and $1.5\sigma$ of their stated statistical uncertainties for NGC\,5024, NGC\,5053, NGC\,5272, NGC\,5466, and NGC\,7099, 
respectively, and well inside the systematic uncertainties for all the clusters, except NGC\,5466.
This may mean that each method of distance determination may have a significant systematics for such distant and RR~Lyrae-contaminated clusters, while a compilation of estimates 
from different methods can provide a much more accurate result.

The same conclusion is derived from comparing parallaxes obtained through various methods. 
We convert the distances and their total uncertainties from our isochrone-fitting into parallaxes and their corresponding uncertainties. 
These are then compared in Table~\ref{parallax} with parallaxes obtained from our analysis and \citetalias{vasiliev2021}, both derived from Gaia DR3 astrometry. 
Additionally, we compare them with parallaxes converted from the \citetalias{baumgardt2021} compiled distances, 
which were obtained using various methods and presented in Table\ref{properties}.
A moderate agreement between the parallaxes is seen.

The total uncertainty of any astrometric estimate of {\it Gaia} DR3 parallax cannot be better than 0.01 mas \citep{vasiliev2021}, 
while the total uncertainty of isochrone-fitting parallax decreases with $R$ (so that the relative parallax uncertainty is constant).
Therefore, Table~\ref{parallax} shows that the parallax estimates from the {\it Gaia} DR3 astrometry are less precise than those from our isochrone fitting for such distant clusters.

\begin{table*}
\def\baselinestretch{1}\normalsize\footnotesize
\caption{The Relative Estimates Presented as Cluster Sequences along Ascending or Descending Parameter. 
}
\label{sequences}
\[
\begin{tabular}{lc}
\hline
\noalign{\smallskip}
Parameter                                           & Sequence \\
\hline
\noalign{\smallskip}
$\Delta$[Fe/H] (dex) from metal-poor to metal-rich  & NGC\,5053 - \textbf{0.02} - NGC\,7099 - \textbf{0.14} - NGC\,5466 - \textbf{0.02} - NGC\,5024 - \textbf{0.35} - NGC\,5272  \\
$\Delta$Age (Gyr) from old to young                 & NGC\,5024 - \textbf{0.00} - NGC\,7099 - \textbf{0.19} - NGC\,5053 - \textbf{0.56} - NGC\,5466 - \textbf{0.63} - NGC\,5272 \\
$\Delta R$ (kpc) from distant to nearby             & NGC\,5024 - \textbf{1.19} - NGC\,5053 - \textbf{1.43} - NGC\,5466 - \textbf{5.45} - NGC\,5272 - \textbf{1.83} - NGC\,7099 \\ 
$\Delta E(B-V)$ (mag) from less to more reddened    & NGC\,5053 - \textbf{0.005} - NGC\,5024 - \textbf{0.000} - NGC\,5272 - \textbf{0.000} - NGC\,5466 - \textbf{0.020} - NGC\,7099 \\
\hline
\end{tabular}
\]
\end{table*}

Comparing our reddening estimates with those in Tables~\ref{properties} and \ref{diversity}, we conclude that 
our estimates are higher than those of \citet{harris} and \citet{schlaflyfinkbeiner2011} for some of the clusters, higher than all estimates of \citet{lallement2019},
agree with the estimates of \citet[][hereafter SFD98]{sfd98}, \citet{planck}, and \citet{dotter2010}, 
while generally lower than the estimates of \citet{green2019}, \citet{gmk2023}, and \citet{paust2010}.

Our final extinction $A_\mathrm{V}$ estimates are $0.08\pm0.01$, $0.06\pm0.01$, $0.08\pm0.01$, $0.08\pm0.01$, and $0.16\pm0.01$\,mag (with the systematic uncertainty 0.03\,mag) for 
NGC\,5024, NGC\,5053, NGC\,5272, NGC\,5466, and NGC\,7099, respectively.

Four of our five clusters are very close to the North Galactic pole. Three of them show $A_\mathrm{V}=0.08$\,mag. Hence, this value can be accepted as a reliable estimate of 
the total Galactic extinction across the whole dust layer above the Sun. Note that the total Galactic extinction below the Sun (i.e. at the South Galactic pole) 
should be few hundredths of a magnitude higher, since the midplane of the Galactic dust layer is below the Sun.

Similar to our previous papers, we consider the relative estimates for the cluster parameters separately derived for each model.
Systematic errors of the models must be canceled out in such relative estimates. 
We use four key CMDs available for all the clusters: (i) $F606W-F814W$ from \citetalias{nardiello2018}, (ii) $G_\mathrm{BP}-G_\mathrm{RP}$ from {\it Gaia} DR3, 
(iii) $B-I$ from \citetalias{stetson2019}, and (iv) $g_\mathrm{PS1}-i_\mathrm{PS1}$ from PS1.
Table~\ref{sequences} presents the relative estimates of the derived parameters as cluster sequences.
Adopting 0.1 dex, 0.4 Gyr, 0.2 kpc, and 0.01 mag as the uncertainties of the relative estimates of [Fe/H], age, distance and reddening, respectively, 
we conclude from Table~\ref{sequences} that
(i) NGC\,5272 is more metal-rich than the remaining clusters of nearly the same [Fe/H]$\approx-2$;
(ii) three clusters have nearly the same old age, while NGC\,5466 is younger and NGC\,5272 is much younger;
(iii) NGC\,7099 has a higher reddening than the remaining clusters with nearly the same very low reddening.
Thus, the most important conclusion is that NGC\,5024, NGC\,5053, and NGC\,7099 have nearly the same metallicity and age. 
Moreover, these clusters have similar low helium enrichments.
Hence, their HB morphology difference should be explained by other parameters besides metallicity, age, or helium enrichment.

\begin{table*}
\def\baselinestretch{1}\normalsize\normalsize
\caption{The Count of the Blue HB, RR~Lyrae and Red HB Stars and HB Type of the Clusters. 
}
\label{hbtype}
\[
\begin{tabular}{llcccc}
\hline
\noalign{\smallskip}
Cluster   &       Data set           & Blue HB & RR Lyrae & Red HB & HB type \\           
\hline
\noalign{\smallskip}
NGC\,5024 & {\it Gaia}               & 188     & 34       & 3      & $0.82\pm0.14$ \\     
NGC\,5024 & \citetalias{stetson2019} & 263     & 39       & 6      & $0.83\pm0.12$ \\     
\noalign{\smallskip}
NGC\,5053 & {\it Gaia}               & 34      & 8        & 1      & $0.77\pm0.18$ \\     
NGC\,5053 & \citetalias{stetson2019} & 35      & 9        & 2      & $0.72\pm0.17$ \\     
\noalign{\smallskip}
NGC\,5272 & {\it Gaia}               & 79      & 91       & 61     & $0.08\pm0.13$ \\     
NGC\,5272 & \citetalias{stetson2019} & 89      & 97       & 61     & $0.11\pm0.13$ \\
\noalign{\smallskip}
NGC\,5466 & {\it Gaia}               & 65      & 19       & 8      & $0.62\pm0.08$ \\     
NGC\,5466 & \citetalias{stetson2019} & 65      & 18       & 8      & $0.63\pm0.08$ \\     
\noalign{\smallskip}
NGC\,7099 & {\it Gaia}               & 82      & 5        & 4      & $0.86\pm0.14$ \\     
NGC\,7099 & \citetalias{stetson2019} & 74      & 5        & 6      & $0.80\pm0.15$ \\
\hline
\end{tabular}
\]
\end{table*}

\subsection{HB Morphology}
\label{hb}

We calculate the HB types of the clusters by use of the {\it Gaia} and \citetalias{stetson2019} data sets, which allow us a robust determination of RR~Lyrae variables by the 
\verb"VarFlag" parameter in the former and \verb"Vary" (Welch-Stetson variability index) and \verb"Weight" (weight of the variability index) parameters in the latter data sets.
Table~\ref{hbtype} shows the count of the blue HB stars, RR~Lyrae variables and red HB stars, used in our HB type calculation.
The derived HB types are similar for different data sets of the same cluster. The HB type uncertainties are calculated by our Monte-Carlo simulation.
The mean values of the HB types are presented in Table~\ref{properties}, where they can be compared with the HB type estimates from \citet{torelli2019} and \citet{arellano2024}. 
A good agreement is seen for NGC\,5024, NGC\,5466, and NGC\,7099.

Our HB type estimates for NGC\,5024, NGC\,5272, and NGC\,7099 may be underestimated due to a strong incompleteness of the {\it Gaia} and \citetalias{stetson2019} data sets 
in crowded centres of these clusters, where many blue HB stars exist \citep{catelan2009}.
Our underestimation of the blue HB star counts in these clusters is evident from a comparison of their counts in Table~2 of \citet{torelli2019} and Table~\ref{hbtype}.
The comparison of these tables also shows that no such an underestimation for loose clusters NGC\,5053 and NGC\,5466, whose fields are well covered by the {\it Gaia} and 
\citetalias{stetson2019} data sets.
However, even a strong underestimation of the blue HB star count by a hundred stars would bias the HB type by only few hundredths, i.e. imperceptibly.

A bias of the HB type estimates of \citet{torelli2019} due to imperfect identification of cluster members and RR~Lyrae variables seems to be more significant.
Indeed, \citet{torelli2019} count 0, 8, and 5 red HB stars in NGC\,5024, NGC\,5053, and NGC\,5466, respectively, in contrast to our counts 6, 2, and 8 presented in Table~\ref{hbtype}.
One can compute that the difference between these counts significantly affects the HB type estimates. 
Therefore, we check properties of all stars, star-by-star, in this sparsely populated red HB domain of these clusters taking into account also red HB stars
from the \citetalias{nardiello2018} data sets covering the central parts of these clusters.
As a result, we conclude that our counts of the red HB stars in NGC\,5024, NGC\,5053, and NGC\,5466 are more reliable, while those of \citet{torelli2019} seem to be wrong.
In particular, NGC\,5466 has more red HB stars than NGC\,5053 (this may indicate that the former is younger than the latter), while most of the eight \citet{torelli2019}'s red HB 
stars in NGC\,5053 are variables or non-members.
This is the reason of a significant difference between our and \citet{torelli2019} HB types for NGC\,5053, while such a difference for NGC\,5272 
is explained by the loss of much more blue than red HB stars in the crowded centre of NGC\,5272 due to the above mentioned strong incompleteness of the {\it Gaia} and 
\citetalias{stetson2019} data sets there.
We conclude that the HB type of NGC\,5053 must be rather close to those of NGC\,5024 and NGC\,7099.\footnote{$\Delta(V-I)$, $\tau_{HB}$ and HB type from Table~\ref{properties}
correlate for these clusters. After our analysis of the NGC\,5053 HB type bias, we propose that the $\Delta(V-I)$ and $\tau_{HB}$ HB morphology indexes of this loose cluster may also 
be biased and should be revised.}
This perfectly agrees with our earlier conclusion that these three clusters have nearly the same [Fe/H], age and helium abundance.
Moreover, our estimate of slightly younger age of NGC\,5466 and much younger age of NGC\,5272 agrees with their lower HB types.
Thus, at first glance, the HB morphology difference of these clusters does not contradict to the suggestion that [Fe/H] and age are the first and second HB parameters.

Table~\ref{hbmass} describes some parameters of the HB star distribution on mass and preceding mass loss.

Regular stellar evolution without an extreme mass loss produces the HB and AGB stars within a certain range of mass, effective temperature, and colour, 
which uniquely correspond to each other: 
mass increases with decreasing temperature and increasing colour from the left to right side of the CMDs in Figs~\ref{gaia}--\ref{ps1}.
The range of mass is predicted, for example, by the best-fitting BaSTI isochrones describing regular stellar evolution for $Y=0.25$ and $0.275$, 
i.e. the red and orange curves, respectively, in Figs~\ref{gaia}--\ref{ps1}.
The maximum HB mass within regular stellar evolution, about 0.8 solar mass, is nearly the same for these low-metallicity and old clusters.
\footnote{To compare the clusters, we show the BaSTI ZAHBs in Figs~\ref{gaia}--\ref{ps1} for nearly the same range of 0.5--0.8 solar mass from 
the left to right side of the ZAHB curve.}
The minimum HB masses within regular stellar evolution $M_{Y=0.25}$ and $M_{Y=0.275}$ for $Y=0.25$ and $0.275$, respectively, are based on the best-fitting BaSTI isochrones, 
averaged for all the key data sets and presented in Table~\ref{hbmass}.
Note that $M_{Y=0.25}>M_{Y=0.275}$.
The observed distribution of the HB stars between the maximum and minimum masses corresponds to a theoretical distribution from BaSTI. 
In particular, Figs~\ref{gaia}--\ref{ps1} show that the oldest and low-metallicity NGC\,5024, NGC\,5053, and NGC\,7099 have much more stars at the blue side 
(i.e. between the areas I and II in terms of Fig.~\ref{hst_details})
of the red and orange BaSTI isochrones in the HB domain than at their red side (i.e. between the areas II and III in terms of Fig.~\ref{hst_details}), 
while younger NGC\,5466 and NGC\,5272 have comparable number of stars at the sides.
Thus, for each cluster, $M_{Y=0.25}$ and $M_{Y=0.275}$ are the unambiguous characteristics of the HB stars within regular stellar evolution.

\begin{table*}
\def\baselinestretch{1}\normalsize\footnotesize
\caption{Some Parameters of the Distribution of the HB Stars on Mass and Preceding Mass Loss. \\
$M_{Y=0.25}$ and $M_{Y=0.275}$ are the minimum HB masses within regular stellar evolution for $Y=0.25$ and $0.275$, respectively,
$M_{Gratton2010}$ is the empirical minimum HB mass estimate from \citet{gratton2010},
$\mu_{1G}$ and $\mu_{2G}$ are the mass loss for the first and second generation stars, respectively, from \citet{tailo2020},
$\eta$ is the parameter in Reimers law of mass loss \citep{reimers} taken from \citet{tailo2020},
$N_{least}$ is the number of observed HB stars with mass lower than $M_{Gratton2010}$,
All masses are in solar mass.
}
\label{hbmass}
\[
\begin{tabular}{lccccccc}
\hline
\noalign{\smallskip}
Cluster   & $M_{Y=0.25}$ & $M_{Y=0.275}$ & $M_{Gratton2010}$ & $\mu_{1G}$ & $\mu_{2G}$ & $\eta$ & $N_{least}$ \\
\hline
\noalign{\smallskip}
NGC\,5024 & $0.670\pm0.005$ & $0.640\pm0.005$ & $0.638\pm0.002$ & $0.100\pm0.017$ & $0.120\pm0.019$ & $0.26\pm0.02$ & $\le15$ \\
NGC\,5053 & $0.700\pm0.005$ & $0.660\pm0.005$ & $0.674\pm0.001$ & $0.116\pm0.014$ &                 & $0.32\pm0.02$ & $\le1$ \\
NGC\,5272 & $0.680\pm0.005$ & $0.670\pm0.005$ & $0.624\pm0.003$ & $0.188\pm0.017$ & $0.240\pm0.022$ & $0.46\pm0.02$ & $\le15$ \\
NGC\,5466 & $0.700\pm0.005$ & $0.670\pm0.005$ & $0.687\pm0.020$ & $0.103\pm0.017$ & $0.119\pm0.023$ & $0.26\pm0.02$ & $\le1$ \\
NGC\,7099 & $0.680\pm0.005$ & $0.650\pm0.005$ & $0.644\pm0.004$ & $0.066\pm0.014$ & $0.083\pm0.019$ & $0.19\pm0.02$ & $\le4$ \\
\hline
\end{tabular}
\]
\end{table*}

There may be HB stars with a mass lower than $M_{Y=0.275}$. 
They are seen in Figs~\ref{gaia}--\ref{ps1} along the BaSTI ZAHB purple and blue curves to the left (i.e. bluer) of the regular BaSTI isochrone red and orange curves.
NGC\,5272 has many such stars,\footnote{This includes all RR~Lyrae variables of this cluster, 
as seen in Figs~\ref{hst} and \ref{hst_details} (the RR~Lyrae variables are eliminated in Figs~\ref{gaia}, \ref{stetson}, and \ref{ps1}).}, while the remaining clusters have only a few.
Accordingly, the empirical minimum HB mass estimates $M_{Gratton2010}$ from \citet{gratton2010}, defined as the values including 90\% of the observed distribution of the HB stars 
and presented in Table~\ref{hbmass}, are close to $M_{Y=0.275}$ for all the clusters, except NGC\,5272 with $M_{Y=0.275}\gg M_{Gratton2010}$.\footnote{Our minimum HB mass 
estimates by use of the best-fitting BaSTI isochrones almost coincide with those of \citet{gratton2010}, which are thus considered hereafter. 
The \citet{gratton2010} estimates are based on an older model. 
Hence, this coincidence indicates that our conclusions about the HB morphology would be the same with any reliable HB model, not only with BaSTI.}
Hence, the HB morphology of NGC\,5272 must be explained either by a very high helium enrichment, which is not observed for this cluster, or an extreme mass loss between the MS and HB,
while the HB morphology of the remaining clusters does not need such an additional parameter.
This agrees with the estimates by \citet{tailo2020} of mass loss for the first ($\mu_{1G}$) and second ($\mu_{2G}$) generation of stars and corresponding parameter $\eta$ 
in Reimers law \citep{reimers} for the first generation stars presented in Table~\ref{hbmass} for all the clusters.
It is seen that the mass-loss efficiency in NGC\,5272 is much higher than in the remaining clusters.
Note that the blue HB stars with a higher $Y\approx0.275$ belong to the second generation of NGC\,5272 with very high mass loss.\footnote{Much longer HB blue tail of 
NGC\,6205, the famous HB second parameter mate of NGC\,5272, can be explained by its even higher mass loss of $0.210\pm0.020$ and $0.273\pm0.021$ solar mass for two generations, 
respectively \citep{tailo2020}.}

The observed distribution of the HB stars on mass often resembles a Gaussian distribution \citep{catelan2009} in accordance with the stochastic mass loss between the MS and HB 
considered by BaSTI.
This suggests a smooth decrease of the HB star count with the decrease of their mass.
Otherwise, an additional parameter is needed in order to explain an abrupt distribution at the minimum HB mass for some clusters.
Such an abrupt distribution can be seen as the absence or very little number of low-mass HB stars to the left of the best-fitting BaSTI isochrone of regular stellar evolution, 
i.e. when $M_{Gratton2010}$ is rather high and $M_{Y=0.275}<M_{Gratton2010}$. Table~\ref{hbmass} shows this for NGC\,5053 and NGC\,5466.
Also, such an abrupt distribution can be seen as a very small number $N_{least}$ (presented in Table~\ref{hbmass}) of observed HB stars with mass lower than $M_{Gratton2010}$ 
(most of these stars are at the centres of the clusters and observed only by the {\it HST}).
Thus, Table~\ref{hbmass} shows that NGC\,5053, NGC\,5466, and NGC\,7099 have rather abrupt distribution at the minimum HB mass.\footnote{Yet another
difference between the clusters is seen in Fig.~\ref{hst}: NGC\,5024 and NGC\,5272 contain much more type II Cepheids, which populate the gap in the middle of the BaSTI AGB,
i.e. the area III in Fig.~\ref{hst_details}.
Type II Cepheids are believed to be the immediate progeny of blue (i.e. low-mass) HB stars \citep{catelan2009}. 
Therefore, the lack of type II Cepheids in NGC\,5053, NGC\,5466, and NGC\,7099 corresponds to the lack of the low-mass HB stars.}
This may be due to the action of one more parameter after metallicity, age, and mass-loss efficiency.
Namely, NGC\,5053, NGC\,5466, and NGC\,7099 have lost their low-mass stars, including the bluest HB stars, due to dynamical evolution and mass segregation.
It seems that both core-collapse and loose clusters lose low-mass stars more effectively than clusters with star concentration in between. 
The low-mass stars are eliminated from the loose clusters NGC\,5053 and NGC\,5466 by Galactic potential, while from NGC\,7099 by its core collapse
\citep{meylan1997,odenkirchen2004,lauchner2006,fekadu2007,beccari2015,kimmig2015,sollima2017,mansfield2022}.
In particular, the large tidal tails of NGC\,5053 and NGC\,5466 suggest that these clusters have been strongly disrupted by interactions with the Galaxy or its satellites.
Thus, it seems that the HB morphology difference of the clusters under consideration can be completely explained by four parameters: 
metallicity, age, mass-loss efficiency, and loss of low-mass stars in cluster evolution.

Note that the pair NGC\,7099 and NGC\,5024 is very similar in its HB morphology difference to the pair NGC\,6397 and NGC\,6809 from \citetalias{ngc6397} as pairs of
a core-collapse and non-core-collapse clusters of similar metallicity, age, helium abundance, and mass-loss efficiency.

\section{Conclusions}
\label{conclusions}

After \citetalias{ngc5904}--\citetalias{ngc6397}, in this study we estimate [Fe/H], age, distance, reddening and extinction of high-latitude low-extinction Galactic globular clusters 
NGC\,5024 (M53), NGC\,5053, NGC\,5272 (M3), NGC\,5466, and NGC\,7099 (M30) 
by fitting BaSTI and DSED theoretical isochrones for [$\alpha$/Fe]$=0.4$ to CMDs based on multiband photometry.
We employed the photometry in, at least, 25 filters from the {\it HST}, {\it Gaia} DR3, PS1, SDSS, SMSS DR3, UKIDSS, VISTA VHS DR5, unWISE, 
large compilation of the $UBVRI$ ground-based observations by \citetalias{stetson2019}, and other data sets, most of which have never been fitted before.
The filters under consideration span a wide wavelength range from the UV to mid-IR. 
{\it HST} and {\it Gaia} DR3 proper motions and parallaxes are used to select the cluster members.
Accordingly, we provided the median parallax and systemic proper motions of the clusters.
Cross-identification of the data sets allowed us to estimate systematic differences between them and verify that the \citetalias{ccm89} extinction law with $R_\mathrm{V}=3.1$
is applicable to these clusters.

The obtained estimates of [Fe/H], age, distance, distance modulus, apparent $V$-band distance modulus, and reddening $E(B-V)$ for all the clusters are presented in Table~\ref{estimates}.
Our estimates of extinction are
$A_\mathrm{V}=0.08$, $0.06$, $0.08$, $0.08$, and $0.16$\,mag for NGC\,5024, NGC\,5053, NGC\,5272, NGC\,5466, and NGC\,7099, respectively, 
with statistic and systematic uncertainties $\pm0.01$ and $\pm0.03$\,mag.
Since three of four clusters near the North Galactic pole demonstrate $A_\mathrm{V}=0.08$, we suggested this value as the total Galactic extinction across the whole Galactic dust
to extragalactic objects at the North polar cap.
Our estimates of all the parameters agree with most estimates from the literature, while disapprove other estimates.
In particular, our [Fe/H] estimates support lower [Fe/H] estimates from the literature for NGC\,5272, while higher [Fe/H] estimates for the remaining low-metallicity clusters.
Accordingly, our estimates support the arguments of \citet{mucciarelli2020} in favour of photometrically and against spectroscopically derived [Fe/H] of clusters with [Fe/H]$\approx-2$.

We recalculated the HB types of these clusters and analysed the distribution of their HB stars on mass. This allowed us to explain their HB morphology difference 
by their different metallicity, age, mass-loss efficiency, and loss of low-mass members, including the bluest HB stars, in the dynamical evolution and mass segregation of 
core-collapse cluster NGC\,7099 and loose clusters NGC\,5053 and NGC\,5466.

NGC\,5024, NGC\,5053, NGC\,5466, and NGC\,7099 have nearly the same metallicity, low helium enrichment, and age (NGC\,5466 may be slightly younger), but only the former retains 
its low-mass stars.
It is worth noting that our results on NGC\,5024, NGC\,5053, NGC\,5466, and NGC\,7099 do not contradict their captured origin from a satellite galaxy suggested by \citet{yoon2002}.

\begin{acknowledgements}
We acknowledge financial support from the Russian Science Foundation (grant no. 20--72--10052).

We thank the anonymous reviewer for useful comments.
We thank
Armando Arellano Ferro for very fruitful discussion of the cluster RR~Lyrae stars,
Giacomo Beccari for providing of a data set with useful comments,
Anupam Bhardwaj for providing of a data set,
Dong-Hwan Cho for providing of a data set,
Sang-Hyun Chun for providing of a data set,
Santi Cassisi for providing the valuable BaSTI isochrones with his exceptionally useful comments,
Aaron Dotter for his comments on DSED,
Heinz Frelijj for providing of a data set,
Gregory Green for discussion of extinction/reddening estimates in the fields of globular clusters,
Frank Grundahl for providing his valuable Str\"omgren data sets with very useful comments,
Young-Beom Jeon for providing of a data set,
K.~J.~Nikitha Jithendran for providing of a data set,
Christopher Onken, Taisia Rahmatulina and Sergey Antonov for their help to access the SkyMapper Southern Sky Survey DR3,
Eric Sandquist for providing of a data set with very useful discussion,
Peter Stetson for providing and having discussion of his valuable $UBVRI$ photometry,
Eugene Vasiliev for his very useful comments on the cluster properties.

This work has made use of BaSTI and DSED web tools;
Filtergraph \citep{filtergraph}, an online data visualization tool developed at Vanderbilt University through the Vanderbilt Initiative in 
Data-intensive Astrophysics (VIDA) and the Frist Center for Autism and Innovation (FCAI, \url{https://filtergraph.com});
the resources of the Centre de Donn\'ees astronomiques de Strasbourg, Strasbourg, France (\url{http://cds.u-strasbg.fr}), including the SIMBAD database, 
the VizieR catalogue access tool \citep{vizier} and the X-Match service;
observations made with the NASA/ESA {\it Hubble Space Telescope};
data products from the {\it Wide-field Infrared Survey Explorer}, which is a joint project of the University of California, Los Angeles, and the Jet Propulsion 
Laboratory/California Institute of Technology;
data products from the Pan-STARRS Surveys (PS1);
data products from the Sloan Digital Sky Survey;
data products from the SkyMapper Southern Sky Survey, SkyMapper is owned and operated by The Australian National University's Research School of Astronomy and Astrophysics,
the SkyMapper survey data were processed and provided by the SkyMapper Team at ANU, the SkyMapper node of the All-Sky Virtual Observatory (ASVO) is hosted at the National 
Computational Infrastructure (NCI);
data products from the Two Micron All Sky Survey, which is a joint project of the University of Massachusetts and the Infrared Processing and Analysis Center/California 
Institute of Technology, funded by the National Aeronautics and Space Administration and the National Science Foundation;
data from the European Space Agency (ESA) mission {\it Gaia} (\url{https://www.cosmos.esa.int/gaia}), processed by the {\it Gaia} Data Processing and Analysis Consortium 
(DPAC, \url{https://www.cosmos.esa.int/web/gaia/dpac/consortium}), and {\it Gaia} archive website (\url{https://archives.esac.esa.int/gaia}).
\end{acknowledgements}

\bibliographystyle{raa}
\bibliography{bibtex}

\begin{thebibliography}{99}


\bibitem[\protect\citeauthoryear{An et al.}{2008}]{an2008} An~D. et al., 2008, \apjs, 179, 326

\bibitem[\protect\citeauthoryear{An et al.}{2009}]{an2009} An~D. et al., 2009, \apj, 700, 523

\bibitem[\protect\citeauthoryear{Anderson et al.}{2008}]{anderson2008} Anderson~J. et al., 2008, \aj, 135, 2055

\bibitem[\protect\citeauthoryear{Arellano Ferro, Giridhar \& Bramich}{2010}]{arellano2010} Arellano~Ferro~A., Giridhar~S., Bramich~D.~M., 2010, \mnras, 402, 226

\bibitem[\protect\citeauthoryear{Arellano Ferro}{2022}]{arellano2022}  Arellano~Ferro~A., 2022, Revista Mexicana de Astronom\'ia y Astrof\'isica, 58, 257

\bibitem[\protect\citeauthoryear{Arellano Ferro}{2024}]{arellano2024}  Arellano~Ferro~A., 2024, in de Grijs~R., Whitelock~P.~A., Catelan~M., eds, 
Proceedings of the International Astronomical Union Vol. 376, At the crossroads of astrophysics and cosmology: Period-luminosity relations in the 2020s. p. 222

\bibitem[\protect\citeauthoryear{Baumgardt \& Vasiliev}{2021}]{baumgardt2021} Baumgardt~H., Vasiliev~E., 2021, \mnras, 505, 5957 (BV21)

\bibitem[\protect\citeauthoryear{Beccari et al.}{2013}]{beccari2013} Beccari~G., Dalessandro~E., Lanzoni~B., Ferraro~F.~R., Sollima~A., Bellazzini~M., Miocchi~P., 2013, \apj, 776, 60

\bibitem[\protect\citeauthoryear{Beccari et al.}{2015}]{beccari2015} Beccari~G., Dalessandro~E., Lanzoni~B., Ferraro~F.~R., Bellazzini~M., Sollima~A., 2015, \apj, 814, 144

\bibitem[\protect\citeauthoryear{Bernard et al.}{2014}]{bernard2014} Bernard~E.~J. et al., 2014, \mnras, 442, 2999

\bibitem[\protect\citeauthoryear{Bica et al.}{2019}]{bica2019} Bica~E., Pavani~D.~B., Bonatto~C.~J., Lima~E.~F., 2019, \aj, 157, 12

\bibitem[\protect\citeauthoryear{Boberg, Friel \& Vesperini}{2015}]{boberg2015} Boberg~O.~M., Friel~E.~D., Vesperini~E., 2015, \apj, 804, 109

\bibitem[\protect\citeauthoryear{Boberg, Friel \& Vesperini}{2016}]{boberg2016} Boberg~O.~M., Friel~E.~D., Vesperini~E., 2016, \apj, 824, 5

\bibitem[\protect\citeauthoryear{Bonatto, Campos \& Kepler}{2013}]{bonatto2013} Bonatto~C., Campos~F., Kepler~S.~O., 2013, \mnras, 435, 263 (BCK13)

\bibitem[\protect\citeauthoryear{Buonanno et al.}{1994}]{buonanno1994} Buonanno~R., Corsi~C.~E., Buzzoni~A., Cacciari~C., Ferraro~F.~R., Fusi~Pecci~F., 1994, \aap, 290, 69

\bibitem[\protect\citeauthoryear{Burger et al.}{2013}]{filtergraph} Burger~D., Stassun~K.~G., Pepper~J., Siverd~R.~J., Paegert~M., De Lee~N.~M., Robinson~W.~H., 2013, 
Astron. Comput., 2, 40

\bibitem[\protect\citeauthoryear{Cardelli, Clayton \& Mathis}{1989}]{ccm89} Cardelli~J.~A., Clayton~G.~C., Mathis~J.~S., 1989, \apj, 345, 245 (CCM89)

\bibitem[\protect\citeauthoryear{Carretta et al.}{2009}]{carretta2009} Carretta~E., Bragaglia~A., Gratton~R., D'Orazi~V., Lucatello~S., 2009, \aap, 508, 695

\bibitem[\protect\citeauthoryear{Carretta et al.}{2010}]{carretta2010} Carretta~E., Bragaglia~A., Gratton~R.~G., Recio-Blanco~A., Lucatello~S., D'Orazi~V., Cassisi~S., 2010, 
\aap, 516, A55

\bibitem[\protect\citeauthoryear{Casagrande \& VandenBerg}{2014}]{casagrande2014} Casagrande~L., VandenBerg~Don~A., 2014, \mnras, 444, 392

\bibitem[\protect\citeauthoryear{Casagrande \& VandenBerg}{2018a}]{casagrande2018a} Casagrande~L., VandenBerg~Don~A., 2018a, \mnras, 475, 5023

\bibitem[\protect\citeauthoryear{Casagrande \& VandenBerg}{2018b}]{casagrande2018b} Casagrande~L., VandenBerg~Don~A., 2018b, \mnras, 479, L102

\bibitem[\protect\citeauthoryear{Catelan}{2009}]{catelan2009} Catelan~M., 2009a, \apss, 320, 261

\bibitem[\protect\citeauthoryear{Chambers et al.}{2016}]{chambers2016} Chambers~K.~C. et al., 2016, arXiv:1612.05560

\bibitem[\protect\citeauthoryear{Chun et al.}{2010}]{chun2010} Chun~S.-H. et al., 2010, \aj, 139, 606 

\bibitem[\protect\citeauthoryear{Chun, Lee \& Lim}{2020}]{chun2020} Chun~S.-H., Lee~J.-J., Lim~D., 2020, \apj, 900, 146

\bibitem[\protect\citeauthoryear{Cohen et al.}{2015}]{cohen2015} Cohen~R.~E., Hempel~M., Mauro~F., Geisler~D., Alonso-Garcia~J., Kinemuchi~K., 2015, \aj, 150, 176

\bibitem[\protect\citeauthoryear{Dalessandro et al.}{2013}]{dalessandro2013} Dalessandro~E., Salaris~M., Ferraro~F.~R., Mucciarelli~A., Cassisi~S., 2013, \mnras, 430, 459

\bibitem[\protect\citeauthoryear{Denissenkov et al.}{2017}]{denissenkov2017} Denissenkov~P.~A., VandenBerg~D.~A., Kopacki~G., Ferguson~J.~W., 2017, \apj, 849, 159

\bibitem[\protect\citeauthoryear{Dotter et al.}{2007}]{dotter2007} Dotter~A., Chaboyer~B., Jevremovi\'c~D., Baron~E., Ferguson~J.~W., Sarajedini~A., Anderson~J., 2007, \aj, 134, 376

\bibitem[\protect\citeauthoryear{Dotter et al.}{2008}]{dotter2008} Dotter~A., Chaboyer~B., Jevremovi\'c~D., Kostov~V., Baron~E., Ferguson~J.W., 2008, \apjs, 178, 89

\bibitem[\protect\citeauthoryear{Dotter et al.}{2010}]{dotter2010} Dotter~A. et al., 2010, \apj, 708, 698

\bibitem[\protect\citeauthoryear{Eisenstein et al.}{2006}]{eisenstein2006} Eisenstein~D.~J. et al., 2006, \apjs, 167, 40

\bibitem[\protect\citeauthoryear{Fekadu et al.}{2007}]{fekadu2007} Fekadu~N., Sandquist~E.~L., Bolte~M., 2007, \apj, 663, 277

\bibitem[\protect\citeauthoryear{Forbes \& Bridges}{2010}]{forbes2010} Forbes~D.~A., Bridges~T., 2010, \mnras, 404, 1203

\bibitem[\protect\citeauthoryear{Goldsbury et al.}{2010}]{goldsbury2010} Goldsbury~R., Richer~H.~B., Anderson~J., Dotter~A., Sarajedini~A., Woodley~K., 2010, \aj, 140, 1830

\bibitem[\protect\citeauthoryear{Gontcharov, Mosenkov \& Khovritchev}{2019}]{ngc5904} Gontcharov~G.~A., Mosenkov~A.~V., Khovritchev~M.~Yu., 2019, \mnras, 483, 4949 (Paper I)

\bibitem[\protect\citeauthoryear{Gontcharov, Khovritchev \& Mosenkov}{2020}]{ngc6205} Gontcharov~G.~A., Mosenkov~A.~V., Khovritchev~M.~Yu., 2020, \mnras, 497, 3674 (Paper II)

\bibitem[\protect\citeauthoryear{Gontcharov et al.}{2021}]{ngc288} Gontcharov~G.~A. et al., 2021, \mnras, 508, 2688 (Paper III)

\bibitem[\protect\citeauthoryear{Gontcharov et al.}{2023a}]{ngc6362} Gontcharov~G.~A. et al., 2023a, \mnras, 518, 3036 (Paper IV)

\bibitem[\protect\citeauthoryear{Gontcharov et al.}{2023b}]{ngc6397} Gontcharov~G.~A. et al., 2023b, \mnras, 526, 5628 (Paper V)

\bibitem[\protect\citeauthoryear{Gontcharov et al.}{2023c}]{gmk2023} Gontcharov~G.~A. et al., 2023c, Astron. Lett., 49, 673

\bibitem[\protect\citeauthoryear{Gratton et al.}{2010}]{gratton2010} Gratton~R.~G., Carretta~E., Bragaglia~A., Lucatello~S., S'Orazii~V., 2010, \aap, 517, A81

\bibitem[\protect\citeauthoryear{Green et al.}{2019}]{green2019} Green~G.~M., Schlafly~E., Zucker~C., Speagle~J.~S., Finkbeiner~D., 2019, \apj, 887, 93 

\bibitem[\protect\citeauthoryear{Grundahl et al.}{1999}]{grundahl1999} Grundahl~F., Catelan~M., Landsman~W.~B., Stetson~P.~B., Andersen~M.~I., 1999, \apj, 524, 242 (GCL99)

\bibitem[\protect\citeauthoryear{Harris}{1996}]{harris} Harris~W.~E., 1996, \aj, 112, 1487

\bibitem[\protect\citeauthoryear{Hewett et al.}{2006}]{ukidss} Hewett~P.~C., Warren~S.~J., Leggett~S.~K., Hodgkin~S.~T., 2006, \mnras, 367, 454

\bibitem[\protect\citeauthoryear{Hidalgo et al.}{2018}]{newbasti} Hidalgo~S.~L. et al., 2018, \apj, 856, 125 

\bibitem[\protect\citeauthoryear{Hunt \& Reffert}{2023}]{hunt2023} Hunt~E.~L., Reffert~S., 2023, \aap, 673, A114

\bibitem[\protect\citeauthoryear{Jeon et al.}{2004}]{jeon2004} Jeon~Y.-B., Lee~M.-G., Kim~S.-L., Lee~H., 2004, \aj, 128, 287

\bibitem[\protect\citeauthoryear{Jurcsik \& Hajdu}{2023}]{jurcsik2023} Jurcsik~J., Hajdu~G., 2023, \mnras, 525, 3486

\bibitem[\protect\citeauthoryear{Kacharov et al.}{2015}]{kacharov2015} Kacharov~N., Koch~A., Caffau~E., Sbordone~L., 2015, \aap, 577, A18

\bibitem[\protect\citeauthoryear{Kains et al.}{2013}]{kains2013} Kains~N. et al., 2013, \aap, 555, A36

\bibitem[\protect\citeauthoryear{Kimmig et al.}{2015}]{kimmig2015} Kimmig~B., Seth~A., Ivans~I.I., Strader~J., Caldwell~N., Anderton~T., Gregersen~D., 2015, \aj, 149, 53

\bibitem[\protect\citeauthoryear{Lallement et al.}{2019}]{lallement2019} Lallement~R., Babusiaux~C., Vergely~J.~L., Katz~D., Arenou~F., Valette~B., Hottier~C., Capitanio~L., 2019, 
\aap, 625, A135

\bibitem[\protect\citeauthoryear{Lauchner, Powell \& Wilhelm}{2006}]{lauchner2006} Lauchner~A., Powell~W.~L., Wilhelm~R., 2006, \apj, 651, L33

\bibitem[\protect\citeauthoryear{Lee, Demarque \& Zinn}{1994}]{lee1994} Lee~Y.-W., Demarque~P., Zinn~R., 1994, \apj, 423, 248

\bibitem[\protect\citeauthoryear{Libralato et al.}{2022}]{libralato2022} Libralato~M. et al., 2022, \apj, 934, 150

\bibitem[\protect\citeauthoryear{Lindegren et al.}{2021}]{lindegren2021} Lindegren~L. et al., 2021, \aap, 649, A4

\bibitem[\protect\citeauthoryear{Mansfield et al.}{2022}]{mansfield2022} Mansfield~S., Dieball~A., Kroupa~P., Knigge~C., Zurek~D.~R., Shara~M., Long~K.~S., 2022, \mnras, 511, 3785

\bibitem[\protect\citeauthoryear{Masseron et al.}{2019}]{masseron2019} Masseron~T. et al., 2019, \aap, 622, A191

\bibitem[\protect\citeauthoryear{Massari et al.}{2016}]{massari2016} Massari~D., Lapenna~E., Bragaglia~A., Dalessandro~E.,  Contreras Ramos~R. Amigo~P., 2016, \mnras, 458, 4162

\bibitem[\protect\citeauthoryear{McMahon et al.}{2013}]{vista} McMahon~R.~G., Banerji~M., Gonzalez~E., Koposov~S.~E., Bejar~V.~J., Lodieu~N., Rebolo~R., VHS Collab., 2013, 
The Messenger, 154, 35

\bibitem[\protect\citeauthoryear{Meisner \& Finkbeiner}{2015}]{planck} Meisner~A.~M., Finkbeiner~D.~P., 2015, \apj, 798, 88

\bibitem[\protect\citeauthoryear{M\'esz\'aros et al.}{2020}]{meszaros2020} M\'esz\'aros~S. et al., 2020, \mnras, 492, 1641

\bibitem[\protect\citeauthoryear{Meylan \& Heggie}{1997}]{meylan1997} Meylan~G., Heggie~D.~C., 1997, \aapr, 8, 1

\bibitem[\protect\citeauthoryear{Milone et al.}{2017}]{milone2017} Milone~A.~P. et al., 2017, \mnras, 464, 3636

\bibitem[\protect\citeauthoryear{Milone et al.}{2018}]{milone2018} Milone~A.~P. et al., 2018, \mnras, 481, 5098

\bibitem[\protect\citeauthoryear{Mucciarelli et al.}{2014}]{mucciarelli2014} Mucciarelli~A., Lovisi~L., Lanzoni~B., Ferraro~F.~R., 2014, \apj, 786, 14

\bibitem[\protect\citeauthoryear{Mucciarelli \& Bonifacio}{2020}]{mucciarelli2020} Mucciarelli~A., Bonifacio~P., 2020, \aap, 640, A87

\bibitem[\protect\citeauthoryear{Nardiello et al.}{2018}]{nardiello2018} Nardiello~D. et al., 2018, \mnras, 481, 3382 (NLP18)

\bibitem[\protect\citeauthoryear{Nikitha, Vig \& Ghosh}{2022}]{nikitha2022} Nikitha~K.~J., Vig~S., Ghosh~S.~K., 2022, \mnras, 514, 5570

\bibitem[\protect\citeauthoryear{Ochsenbein, Bauer \& Marcout}{2000}]{vizier} Ochsenbein~F., Bauer~P., Marcout~J., 2000, \aaps, 143, 23

\bibitem[\protect\citeauthoryear{Odenkirchen \& Grebel}{2004}]{odenkirchen2004} Odenkirchen~M., Grebel~E.~K., 2004, ASP Conference Series, 327, 284

\bibitem[\protect\citeauthoryear{Onken et al.}{2019}]{onken2019} Onken~C.~A. et al., 2019, \pasa, 36, 33

\bibitem[\protect\citeauthoryear{Paust et al.}{2010}]{paust2010} Paust~N.~E.~Q. et al., 2010, \aj, 139, 476

\bibitem[\protect\citeauthoryear{Pietrinferni et al.}{2021}]{pietrinferni2021} Pietrinferni~A. et al., 2021, \apj, 908, 102

\bibitem[\protect\citeauthoryear{Piotto et al.}{2002}]{piotto2002} Piotto~G. et al., 2002, \aap, 391, 945

\bibitem[\protect\citeauthoryear{Piotto et al.}{2015}]{piotto2015} Piotto~G. et al., 2015, \aj, 149, 91

\bibitem[\protect\citeauthoryear{Preet Kaur \& Joshi}{2022}]{preetkaur2022} Preet~Kaur~K., Joshi~P.~S., 2022, arXiv:2209.03019

\bibitem[\protect\citeauthoryear{Reimers}{1975}]{reimers} Reimers~D., 1975, Mem. Soc. R. Sci. Liege, 8, 369

\bibitem[\protect\citeauthoryear{Rey et al.}{1998}]{rey1998} Rey~S.-C., Lee~Y.-W., Byun~Y.-I., Chun~M.-S., 1998, \aj, 116, 1775

\bibitem[\protect\citeauthoryear{Rey et al.}{2001}]{rey2001} Rey~S.-C., Yoon~S.-J., Lee~Y.-W., Chaboyer~B., Sarajedini~A., 2001, \aj, 122, 3219

\bibitem[\protect\citeauthoryear{Riello et al.}{2021}]{riello2021} Riello~M. et al., 2021, \aap, 649, A3

\bibitem[\protect\citeauthoryear{Sandage}{1953}]{sandage1953} Sandage~A.~R., 1953, \aj, 58, 61

\bibitem[\protect\citeauthoryear{Sandquist et al.}{1999}]{sandquist1999} Sandquist~E.~L., Bolte~M., Langer~G.~E., Hesser~J.~E., Mendes~de~Oliveira~C., 1999, \apj, 518, 262

\bibitem[\protect\citeauthoryear{Sarajedini \& Milone}{1995}]{sarajedini1995} Sarajedini~A., Milone~A.~A.~E., 1995, \aj, 109, 269 

\bibitem[\protect\citeauthoryear{Schlafly \& Finkbeiner}{2011}]{schlaflyfinkbeiner2011} Schlafly~E.~F., Finkbeiner~D.~P., 2011, \apj, 737, 103

\bibitem[\protect\citeauthoryear{Schlafly et al.}{2019}]{unwise} Schlafly~E.~F., Meisner~A.~M., Green~G.~M., 2019, \apjs, 240, 30 

\bibitem[\protect\citeauthoryear{Schlegel, Finkbeiner \& Davis}{1998}]{sfd98} Schlegel~D.~J., Finkbeiner~D.~P., Davis~M., 1998, \apj, 500, 525 (SFD98)

\bibitem[\protect\citeauthoryear{Simioni et al.}{2018}]{simioni2018} Simioni~M. et al., 2018, \mnras, 476, 271

\bibitem[\protect\citeauthoryear{Skrutskie et al.}{2006}]{2mass} Skrutskie~M.F. et al., 2006, \aj, 131, 1163

\bibitem[\protect\citeauthoryear{Sollima et al.}{2017}]{sollima2017} Sollima~A., Dalessandro~E., Beccari~G., Pallanca~C., 2017, \mnras, 464, 3871

\bibitem[\protect\citeauthoryear{Stenning et al.}{2016}]{stenning2016} Stenning~D.~C., Wagner-Kaiser~R., Robinson~E., van~Dyk~D.~A., von~Hippel~T., Sarajedini~A., Stein~N., 2016,
\apj, 826, 41

\bibitem[\protect\citeauthoryear{Stetson et al.}{2019}]{stetson2019} Stetson~P.~B., Pancino~E., Zocchi~A., Sanna~N., Monelli~M., 2019, \mnras, 485, 3042 (SPZ19)

\bibitem[\protect\citeauthoryear{Tailo et al.}{2020}]{tailo2020} Tailo~M. et al., 2020, \mnras, 498, 5745

\bibitem[\protect\citeauthoryear{Torelli et al.}{2019}]{torelli2019} Torelli~M. et al., 2019, \aap, 629, A53

\bibitem[\protect\citeauthoryear{Valcarce et al.}{2016}]{valcarce2016} Valcarce~A.~A.~R., Catelan~M., Alonso-Garc\'ia~J., Contreras~Ramos~R., Alves~S., 2016, \aap, 589, A126

\bibitem[\protect\citeauthoryear{Valcin et al.}{2020}]{valcin2020} Valcin~D., Bernal~J.~L., Jimenez~R., Verde~L., Wandelt~B.~D., 2020, Journal of Cosmology and Astroparticle Physics, 
12, 2

\bibitem[\protect\citeauthoryear{Vallenari et al.}{2023}]{gaiadr3} Vallenari~A. et al., 2023, \aap, 674, A1

\bibitem[\protect\citeauthoryear{VandenBerg et al.}{2013}]{vandenberg2013} VandenBerg~Don~A., Brogaard~K., Leaman~R., Casagrande~L., 2013, \apj, 775, 134

\bibitem[\protect\citeauthoryear{Vasiliev \& Baumgardt}{2021}]{vasiliev2021} Vasiliev~E., Baumgardt~H., 2021, \mnras, 505, 5978 (VB21)

\bibitem[\protect\citeauthoryear{Vitral}{2021}]{vitral2021} Vitral~E., 2021, \mnras, 504, 1355

\bibitem[\protect\citeauthoryear{Wright et al.}{2010}]{wise} Wright~E.~L. et al., 2010, \aj, 140, 1868 

\bibitem[\protect\citeauthoryear{Yang et al.}{2023}]{yang2023} Yang~Y., Zhao~J.-K., Tang~X.-Z., Ye~X.-H., Zhao~G., 2023, \apj, 953, 130

\bibitem[\protect\citeauthoryear{Yoon \& Lee}{2002}]{yoon2002} Yoon~S.-J., Lee~Y.-W., 2002, \sci, 297, 578

\end{thebibliography}

\appendix

\section{The other key CMDs of the clusters}
\label{addcmds}

\begin{figure}
\includegraphics{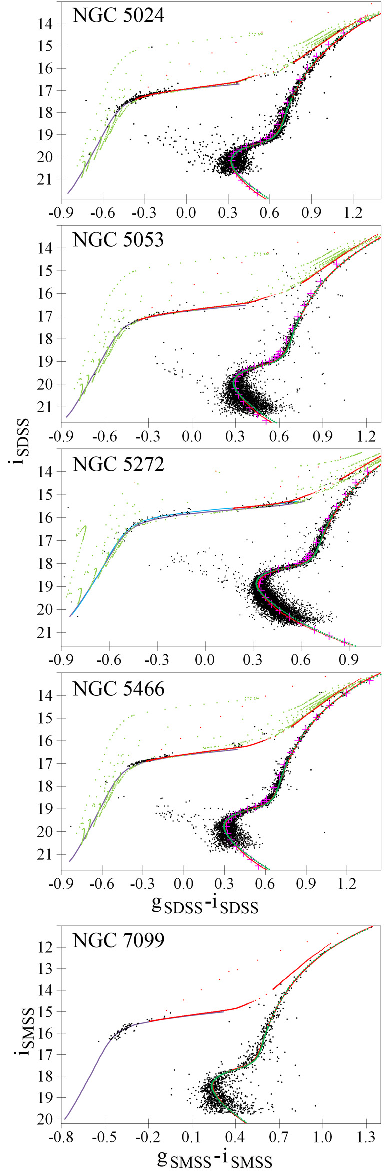}
\caption{$g_\mathrm{SDSS}-i_\mathrm{SDSS}$ versus $i_\mathrm{SDSS}$ CMDs for the {\it Gaia} cluster members from the SDSS data sets for NGC\,5025, NGC\,5272, NGC\,5466
and all the stars from the initial SDSS data set for NGC\,5053, 
as well as $g_\mathrm{SMSS}-i_\mathrm{SMSS}$ versus $i_\mathrm{SMSS}$ CMD for the {\it Gaia} cluster members from the SMSS data set for NGC\,7099.
The fiducial sequences of \citet{an2008} are shown as the magenta crosses.
The isochrones for $Y=0.25$ from BaSTI (red), DSED (green), BaSTI ZAHB (purple), DSED HB/AGB (light green), as well as the NGC\,5272 BaSTI ZAHB for $Y=0.275$ (blue) are calculated with 
the best-fitting parameters from Table~\ref{cmds}.
RR~Lyrae variables are eliminated, except the plot for NGC\,5053.
}
\label{sdsssmss}
\end{figure}

\begin{figure}
\includegraphics{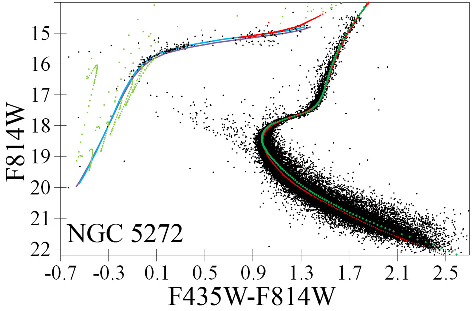}
\caption{{\it HST}/ACS $F435W-F814W$ versus $F814W$ CMD for the NGC\,5272 \citet{libralato2022} data set.
The isochrones for $Y=0.25$ from BaSTI (red), DSED (green), BaSTI ZAHB (purple), and DSED HB/AGB (light green) are calculated with the best-fitting parameters from Table~\ref{cmds}.
RR~Lyrae variables are retained.
}
\label{libralato}
\end{figure}

\begin{figure}
\includegraphics{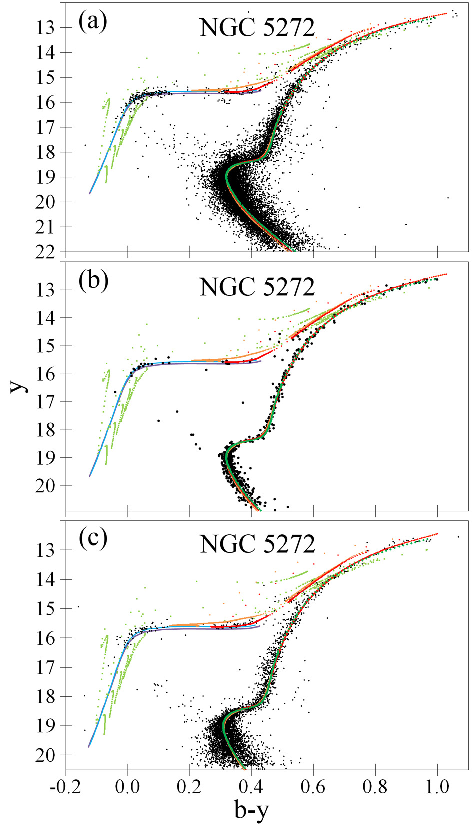}
\caption{Str\"omgren $b-y$ versus $y$ CMDs for NGC\,5272:
(a) all the stars from the \citetalias{grundahl1999} data set,
(b) the {\it Gaia} cluster members from the \citetalias{grundahl1999} data set,
(c) the {\it Gaia} cluster members from the \citet{massari2016} data set.
The isochrones for $Y=0.25$ from BaSTI (red), DSED (green), BaSTI ZAHB (purple), DSED HB/AGB (light green), as well as for $Y=0.275$ from BaSTI (orange) and BaSTI ZAHB (blue) 
are calculated with the best-fitting parameters from Table~\ref{cmds}.
RR~Lyrae variables are retained in the plot (a), while eliminated in the plots (b) and (c).
}
\label{stromgren}
\end{figure}

\label{lastpage}
\end{document}